\documentclass[iop]{emulateapj}

\shorttitle{X-ray Spectral and Optical Properties of a ULX}
\shortauthors{Avdan et al.}

\begin{document}


\title{X-RAY SPECTRAL AND OPTICAL PROPERTIES OF A ULX IN NGC 4258 (M106)}


\author{H. Avdan\altaffilmark{1,2}, S. Avdan\altaffilmark{1,2}, A. Akyuz\altaffilmark{1,2}, S. Balman\altaffilmark{3}, N. Aksaker\altaffilmark{2,4}, I. Akkaya Oralhan\altaffilmark{5}}

\email{avdan.hsn@gmail.com}


\altaffiltext{1}{Department of Physics, Cukurova University, 01330 Adana, Turkey}
\altaffiltext{2}{Space Sciences and Solar Energy Research and Application Center (UZAYMER), Cukurova University, 01330 Adana, Turkey}
\altaffiltext{3}{Department of Physics, Middle East Technical University, 06800 Ankara, Turkey}
\altaffiltext{4}{Vocational School of Technical Sciences, Cukurova University, 01410 Adana, Turkey}
\altaffiltext{5}{Department of Astronomy and Space Sciences, Erciyes University, 38039 Kayseri, Turkey}


\begin{abstract}

We study the X-ray and optical properties of the ultraluminous X-ray source (ULX) X-6 in the nearby galaxy NGC 4258 (M106) based on the archival {\it XMM-Newton}, {\it Chandra}, {\it Swift}, and {\it Hubble Space Telescope} ({\it HST}) observations. The source has a peak luminosity of $L_{\mathrm{X}} \sim 2 \times 10^{39}$ erg s$^{-1}$ in the {\it XMM-Newton} observation of 2004 June. Consideration of the hardness ratios and spectral model parameters shows that the source seems to exhibit possible spectral variations throughout the X-ray observations. In the images from the {\it HST}/Advanced Camera for Surveys (ACS), three optical sources have been identified as counterpart candidates within the 1$\sigma$ error radius of 0$\arcsec$.3. The brightest one has an absolute magnitude of $M_{V} \approx$ $-$7.0 and shows extended structure. The remaining two sources have absolute magnitudes of $M_{V} \approx$ $-$5.8 and $-$5.3 mag. The possible spectral types of the candidates from brightest to dimmest were determined as B6$-$A5, B0$-$A7, and B2$-$A3, respectively. The counterparts of the X-ray source possibly belong to a young star cluster. Neither the standard disk model nor the slim disk model provides firm evidence to determine the spectral characteristics of ULX X-6. We argue that the mass of the compact object lies in the range $10-15M_{\sun}$ indicating that the compact source is most likely a stellar-mass black hole.  

\end{abstract}





\section{Introduction}

Ultraluminous X-ray sources (ULXs) are extragalactic off-nuclear point-like sources with luminosities exceeding the Eddington limit for a 10 $M_{\sun}$ black hole (BH) ($L_{\mathrm{X}}$ $\textgreater$ $10^{39}$ erg s$^{-1}$) \citep{fen11}. If the emission is isotropic, the compact objects in some of the bright ULXs might be intermediate-mass BHs with masses $\sim$ 10$^{2}-$10$^{4}$ $M_{\sun}$ \citep{mil04}. Conversely, some ULXs might contain stellar-mass BHs and their high luminosities may arise from supercritical accretion \citep{sha73,pou07}. Recent studies on ULXs showed that stellar-mass BH scenarios are reliable \citep{liu13,mot13,fab15}. On the other hand, pulsations with an average period of 1.37 s were detected from a ULX in M82 using {\it NuSTAR} data, which indicates that the compact object in this system is a neutron star \citep{bac14}. That result has led to the idea that some ULX systems may harbour neutron stars instead of BHs. The nature of the ULX binary systems is still unclear.

Studying the X-ray spectral states and state transitions of ULXs with the help of available multi-epoch data and comparing them with the well-known characteristics of Galactic BH binaries (BHB) are essential tools for understanding the radiative mechanisms of these sources. There are three active states that have been defined for Galactic BHBs: thermal, hard, and steep power law (PL). In the thermal state, a geometrically thin, optically thick accretion disk dominates the emission, while in the hard state the emission is produced by a geometrically thick, optically thin Comptonizing region. The hard state is characterized by non-thermal PL emission with a photon index $1.4<\Gamma<2.1$. However, the steep PL state is defined by a softer spectrum having a photon index of $\Gamma>2.4$ \citep{rem06}. In the steep PL or thermal state, most of the Galactic BHBs have relatively higher luminosities than in the hard state. A similar correlation between luminosity and photon index has been found in ULXs X-1 in NGC 1313 \citep{fen06,dew10} and X37.8+54 in M82 \citep{jin10}, although there are some ULXs that exhibit the opposite behavior (NGC 1313 X-2, \citealt{fen06}; NGC 4736 X-2, \citealt{avd14}). Additionally, distinct spectral state transitions have been observed in some ULXs (e.g. NGC 2403 src 3, \citealt{iso09}; IC 342 X-1, \citealt{mar14}).

On the other hand, identification of the optical counterparts of the ULXs may provide valuable information. The optical emission could originate from the donor star and/or the accretion disk via X-ray photoionization \citep{fen11}. The optical counterparts of the several ULXs have been found in nearby galaxies using {\it Hubble Space Telescope} ({\it HST}) data (\citealt{tao11} and references therein; \citealt{gla13}). Broadband {\it HST} photometry of the optical counterparts allows constraints to be placed on the mass and spectral type of the companion star \citep{yan11,gri11,gri12}. These constraints could also be defined by studying the environment of the ULX if the system belongs to a stellar cluster or association \citep{gri11,pou13}.

In this work, the X-ray spectral properties of the ULX X-6 \footnote{We adopted the source number from the work of Akyuz et al. (2013). They numbered the detected sources in NGC 4258 as XMM-$n$, where $n$ represented the source number with decreasing EPIC pn count rate. We have shortened their designation to X-6 for convenience.} in NGC 4258 have been studied using archival {\it XMM-Newton}, {\it Chandra} and {\it Swift} observations. Also the optical counterpart of X-6 has been searched on the {\it HST}/Advanced Camera for Surveys (ACS)/WFC archival images. NGC 4258 (M106) is a nearby (7.7 Mpc, \citealt{swa11}) Seyfert-type spiral galaxy. It is well known for its anomalous arms, discovered on the basis of H$\alpha$ imaging \citep{wil01}. The X-6 is located 2$\arcmin$.4 away from the center of the galaxy and its {\it Chandra} coordinate is R.A. $=$ 12$^{\mathrm{h}}$18$^{\mathrm{m}}$43$^{\mathrm{s}}.887$, decl. $=$ +47$^{\circ}$17$\arcmin$31$\arcsec.81$. The source was classified as a ULX by \citet{swa11} with an unabsorbed X-ray luminosity of 1.6$\times 10^{39}$ erg s$^{-1}$ in the 0.3$-$10 keV energy band. X-6 is not positionally coincident with any X-ray point source in the {\it Einstein} and {\it ROSAT} catalogs. \citet{aky13} also studied the X-ray spectrum and the temporal properties of this source. They presented spectral and timing analyses based on the {\it XMM-Newton} observations with the longest exposure available for the non-nuclear X-ray point sources in the $D_{25}$ of NGC 4258.

The paper is organized as follows: the observations and data reductions are described in Section 2. The details and results of the analyses are given in Section 3. Discussion of the physical properties of the ULX and a summary are given in Section 4.

\section{Observations}
\subsection{X-ray Data}

NGC 4258 X-6 was observed multiple times with {\it XMM-Newton}, {\it Chandra} and {\it Swift} over 14 years. We have reanalyzed all seven {\it XMM-Newton}, one {\it Chandra} and 12 {\it Swift} observations which are listed in Table 1 with labels, IDs, dates, and good exposures, which indicate exposure times after the removal of background flares. Only observation XM7 was affected by high background flarings, which were excluded from the data (last $\sim$3 ks).

{\it XMM-Newton} data reductions were carried out using the {\scshape sas} (Science Analysis Software, version 13.05).\footnote{http://xmm.esac.esa.int/sas/} {\scshape epchain} and {\scshape emchain} tasks were used to obtain EPIC pn and MOS event files for each observation. The events corresponding to PATTERN$\leq$12 and PATTERN$\leq$4 with FLAG = 0 were selected for EPIC MOS and pn cameras, respectively. The source and background spectra were extracted with the {\scshape evselect} task using appropriate circular regions of 15$\arcsec$. The background regions were selected from a source-free region on the same chip as the source. 

{\it Chandra} data reductions were performed using the {\scshape ciao} (Chandra Interactive Analysis of Observations, version 4.6) software with the {\scshape caldb}package (version 4.5.9).\footnote{http://cxc.harvard.edu/ciao/} The source was located on the ACIS$-$S3 (back-illuminated) chip. We obtained the level 2 event files using {\scshape chandra\_repro} script. The {\scshape specextract} and {\scshape dmextract} tasks were used to generate spectrum and light-curve files using circular regions of 10$\arcsec$. Background photons were extracted by selecting source-free circular regions near the source. 

{\it Swift} XRT data extractions were done with {\scshape xselect} software (version 2.4).\footnote{https://heasarc.gsfc.nasa.gov/ftools/xselect/} The source and background photons were extracted using circular regions of 20$\arcsec$. We detected the source in the majority of the {\it Swift} data sets, but it was not detected in two observations (S3 and S11). The statistical quality of the data was not adequate to perform a spectral analysis.

\begin{deluxetable}{llccc}
\tablewidth{0pt}
\tablecaption{{\it XMM-Newton}, {\it Chandra} and {\it Swift} observations used in this work}
\tablehead{
\colhead{} & \colhead{Label} & \colhead{ObsID} & \colhead{Date} & \colhead{Good Exp.} \\
 & & & & (ks)}
\startdata
{\it XMM-Newton} & XM1 & 0110920101 & 2000 Dec 8 & 16 \\
& XM2 & 0059140101 & 2001 May 6 & 9 \\
& XM3 & 0059140201 & 2001 Jun 17 & 10 \\
& XM4 & 0059140401 & 2001 Dec 17 & 12 \\
& XM5 & 0059140901 & 2002 May 22 & 14 \\ 
& XM6 & 0203270202 & 2004 Jun 1 & 47 \\
& XM7 & 0400560301 & 2006 Nov 17 & 59 \\
\tableline
{\it Chandra}& C1 & 1618 & 2001.05.28 & 21 \\
\tableline
{\it Swift} XRT & S1 & 00037259001 & 2008 Mar 1 & 10 \\
& S2 & 00037317001 & 2008 May 6 & 3 \\
& S3 & 00037317002 & 2009 Mar 9 & 4 \\
& S4 & 00037317003 & 2009 May 9 & 2 \\
& S5 & 00037259002 & 2014 May 21 & 2 \\
& S6 & 00037259005 & 2014 May 24 & 1 \\
& S7 & 00037259006 & 2014 May 25 & 2 \\
& S8 & 00080599001 & 2014 May 25 & 2 \\
& S9 & 00037259007 & 2014 May 30 & 2 \\
& S10 & 00037259009 & 2014 Jun 08 & 2 \\
& S11 & 00037259011 & 2014 Jun 18 & 0.03 \\
& S12 & 00037259012 & 2014 Jun 22 & 2
\enddata
\end{deluxetable}

\subsection{Optical Data}

Observations in the {\it HST}/ACS/WFC data archive were used to look for the optical counterpart of X-6. A summary of {\it HST} observations used in this study is given in Table 2. The three-color optical image of NGC 4258 from the Sloan Digital Sky Survey (SDSS) is shown in Figure 1.

In {\it HST}/ACS/WFC images, the ULX counterpart appears in a star cluster. The relative astrometry between {\it Chandra} and {\it HST} was improved to determine the position of the optical counterpart accurately. C1 data and {\it HST}/ACS/WFC F435W, F555W, and F814W drizzled images were used for astrometric correction. We performed source detection using {\scshape daofind} in {\scshape iraf} for {\it HST} and the {\scshape wavdetect} task in {\scshape ciao} for {\it Chandra}. Then, the sources detected in these images were compared in order to find reference objects to calculate the relative shift between the {\it Chandra} and {\it HST} images. 

We found two appropriate reference sources in the F435W image, but only one in the F555W and F814W images because the other one was out of the frame. Therefore, the F435W image was adopted for astrometric correction by using these two sources as reference objects. One of the sources is the center of the host galaxy and the other one is a point source (R.A. $=$ 12$^{\mathrm{h}}$18$^{\mathrm{m}}$49$^{\mathrm{s}}$.489, decl. $=$ +47$^{\circ}$16$\arcmin$46$\arcsec.55$) on the same chip as X-6. They have $\sim$2860 and 155 counts in C1 data and hence low statistical positional errors of $\sim$0$\arcsec$.02 and $\sim$0$\arcsec$.08, respectively.

After the astrometric correction, the position of the candidate counterpart was derived with a positional error of 0$\arcsec$.3 as R.A. $=$ 12$^{\mathrm{h}}$18$^{\mathrm{m}}$43$^{\mathrm{s}}$.887, decl. $=$ +47$^{\circ}$17$\arcmin$31$\arcsec$.56 on the {\it HST}/ACS/WFC F435W image. The {\it HST}/ACS/WFC F435W, F555W, and F814W images of NGC 4258 together with the corrected position circle on the F435W image are shown in Figure 2. Three possible optical counterparts are identified for X-6 (source 1, 2, and 3) within the error radius (0$\arcsec$.3). All three sources were detected with signal-to-noise ratios (S/Ns) $\ga$ 30. Multiple counterpart candidates have also been reported in NGC 1073 ULX, IC 342 X-1, and M82 X-1 \citep{kaa05,fen08,wan15}.

We have also observed the region of X-6 with the RTT-150 (Russian$-$Turkish Telescope, 150 cm) at TUG (TUBITAK National Observatory, Antalya, Turkey) to confirm that the star cluster belongs to the host galaxy. Using the 1$\arcsec$.7 slit width, spectroscopic data of the region were obtained by TFOSC (Turkish Faint Object Spectrograph and Camera) instrument with grism 15 (dispersion of 3 \AA\ pixel$^{-1}$). A total exposure of 3600 s was acquired. The standard data reduction steps (bias subtraction, flat-field correction, wavelength and flux calibrations) were performed using {\scshape iraf} software (version 2.16.1).\footnote{http://iraf.noao.edu/} Neon lamps were used for wavelength calibration. We calculated the redshift of the region from the observed H$\alpha$ ($\lambda$6563), H$\beta$ ($\lambda$4861), [O {\scshape iii}] ($\lambda$5007 and $\lambda$4959), [S {\scshape ii}] ($\lambda$6717 and $\lambda$6731), and [N {\scshape ii}] ($\lambda$6583) emission lines. The average of the calculated redshifts is $z$ $\sim 0.0016$. This value is similar to the redshift of NGC 4258 ($z_{\mathrm{g}} = 0.001494\pm0.000010$, \citealt{dev91}). This result indicates that the cluster and possible optical counterparts of X-6 may belong to the host galaxy.

\begin{deluxetable}{llccc}
\tablewidth{0pt}
\tablecaption{The log of {\it HST}/ACS observations used in this work}
\tablehead{
\colhead{Filter}\tablenotemark{a} & \colhead{Data Set} & \colhead{Date} & \colhead{Exposure}  \\
 & & & (ks)}
\startdata
F435W & JB1F87JHQ & 2010 May 30 & 0.360 \\
F555W & JB1F87010 & 2010 May 30 & 0.975 \\
F814W & JB1F87JEQ & 2010 May 30 & 0.360 \\
F606W & J96H27020 & 2005 Mar 7 & 1.014 \\
F606W & J96H28020 & 2005 Mar 9 & 1.014 \\
F606W & J96H29020 & 2005 Mar 10 & 1.014 
\enddata
\tablenotetext{a}{The bandwidths of filters are $\lambda$3610$-$$\lambda$4860 \AA\ for F435W, $\lambda$4584$-$$\lambda$6209 \AA\ for F555W, $\lambda$4634$-$$\lambda$7180 \AA\ for F606W, and $\lambda$6885$-$$\lambda$9647 \AA\ for F814W.}
\end{deluxetable}

\begin{figure}
\centering
\includegraphics[scale=0.50]{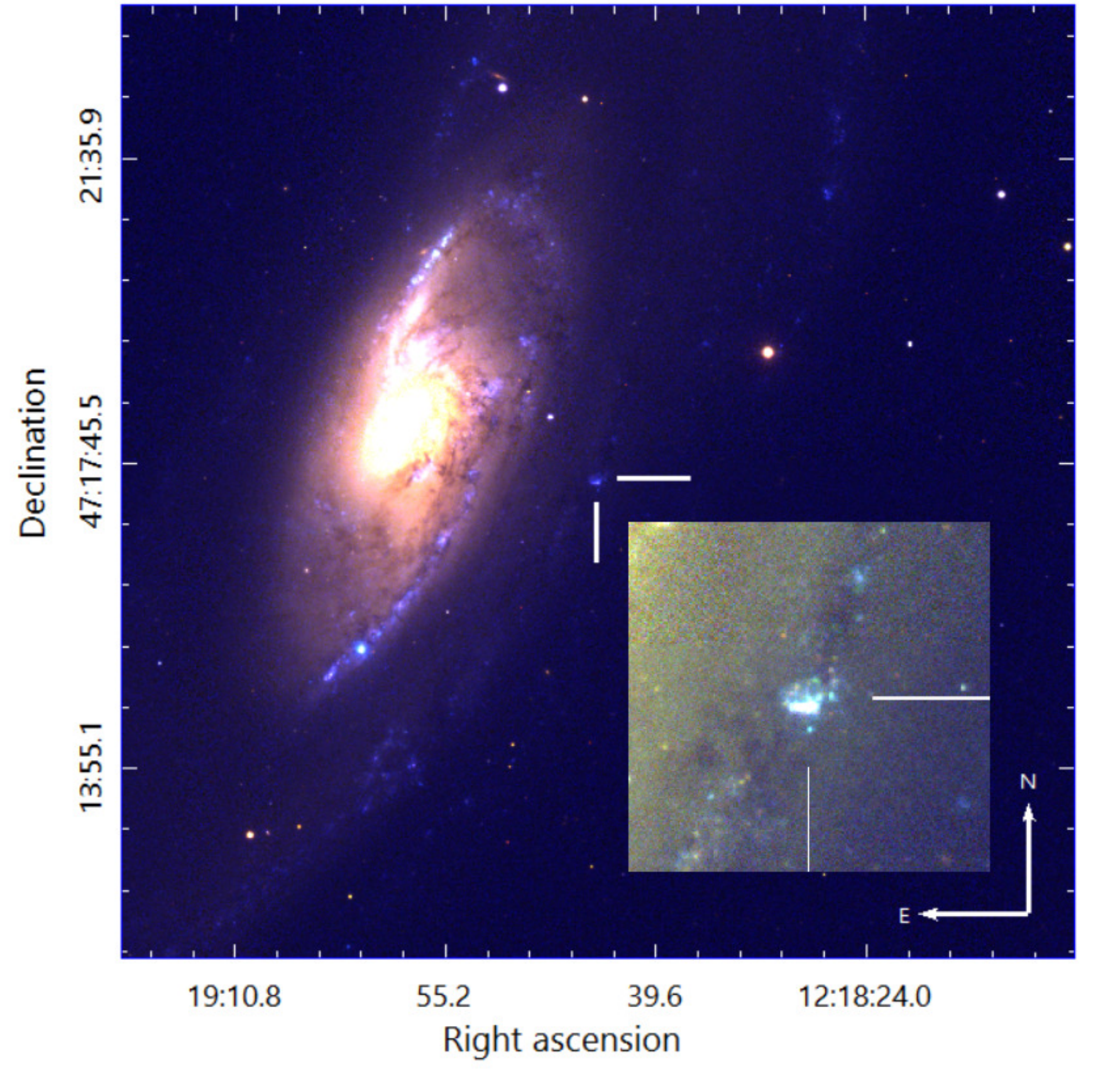}
\caption{The three-color optical SDSS image of NGC 4258. Red, green, and blue colors represent the SDSS i, r, and u bands, respectively. The white lines shows the bright cluster hosting the ULX.}
\end{figure}

\begin{figure*}
\centering
\includegraphics[scale=0.62]{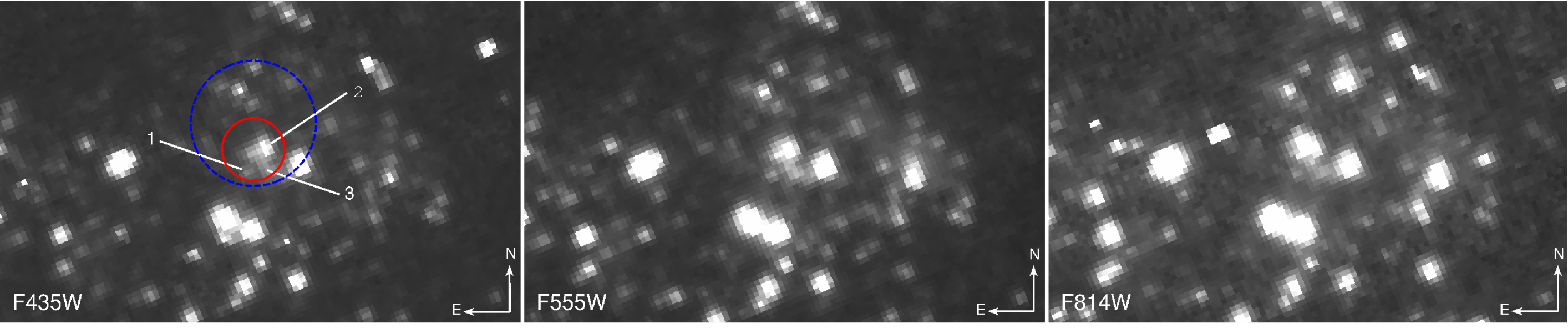}
\caption{{\it HST}/ACS images of the region around NGC 4258 X-6 with three filters. The images have a size of $\sim$ $5\arcsec \times 3\arcsec$. The blue dashed circle represents the original {\it Chandra} position with an accuracy of 0$\arcsec$.6 and the red circle represents the corrected position with an accuracy of 0$\arcsec$.3. Three possible counterparts (source 1$-$3) are found within the error radius.}
\end{figure*}

\section{Data Analysis and Results}
\subsection{X-ray}

Investigation of hardness variability may help to define the states and state transition of the source. Therefore, the events were filtered in three different energy ranges: soft (S) 0.3$-$2 keV, hard (H) 2$-$8 keV, and total 0.3$-$8 keV. Then the net count rates of the ULX were derived for each data set. The {\it XMM-Newton} EPIC pn data were mostly used to calculate the count rates of the source. However, only EPIC MOS data were used for XM4 data because X-6 was partly on the EPIC pn chip gap. To eliminate the differences in sensitivity, the EPIC MOS count rates obtained for XM4 data and the {\it Chandra} count rate from C1 data were converted to {\it XMM-Newton} EPIC pn count rates. The conversions were done with the {\it Chandra} {\scshape pimms} toolkit\footnote{http://cxc.harvard.edu/toolkit/pimms.jsp} by using the best PL parameters calculated for XM4 and C1 data sets. While converting the {\it Chandra} count rates, the Cycle-3 calibration files were used. The light curves obtained from {\it XMM-Newton} and Chandra data are given in Figure 3(a) and those from {\it Swift} in Figure 3(b). 

\begin{figure*}

{
(a)
\label{fig:sub:a}
\includegraphics[scale=0.5]{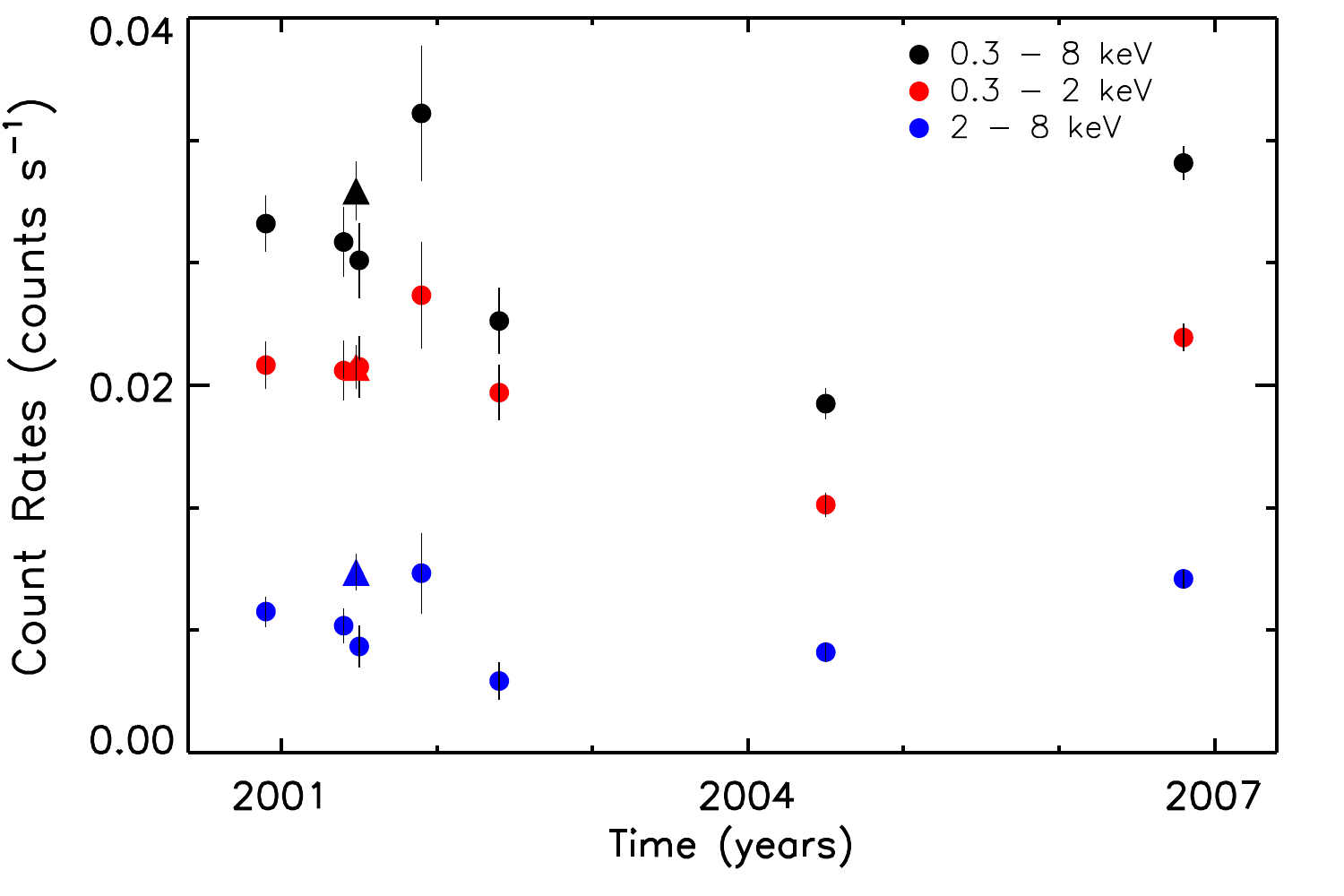}
}
(b)
{
\label{fig:sub:b}
\includegraphics[scale=0.5]{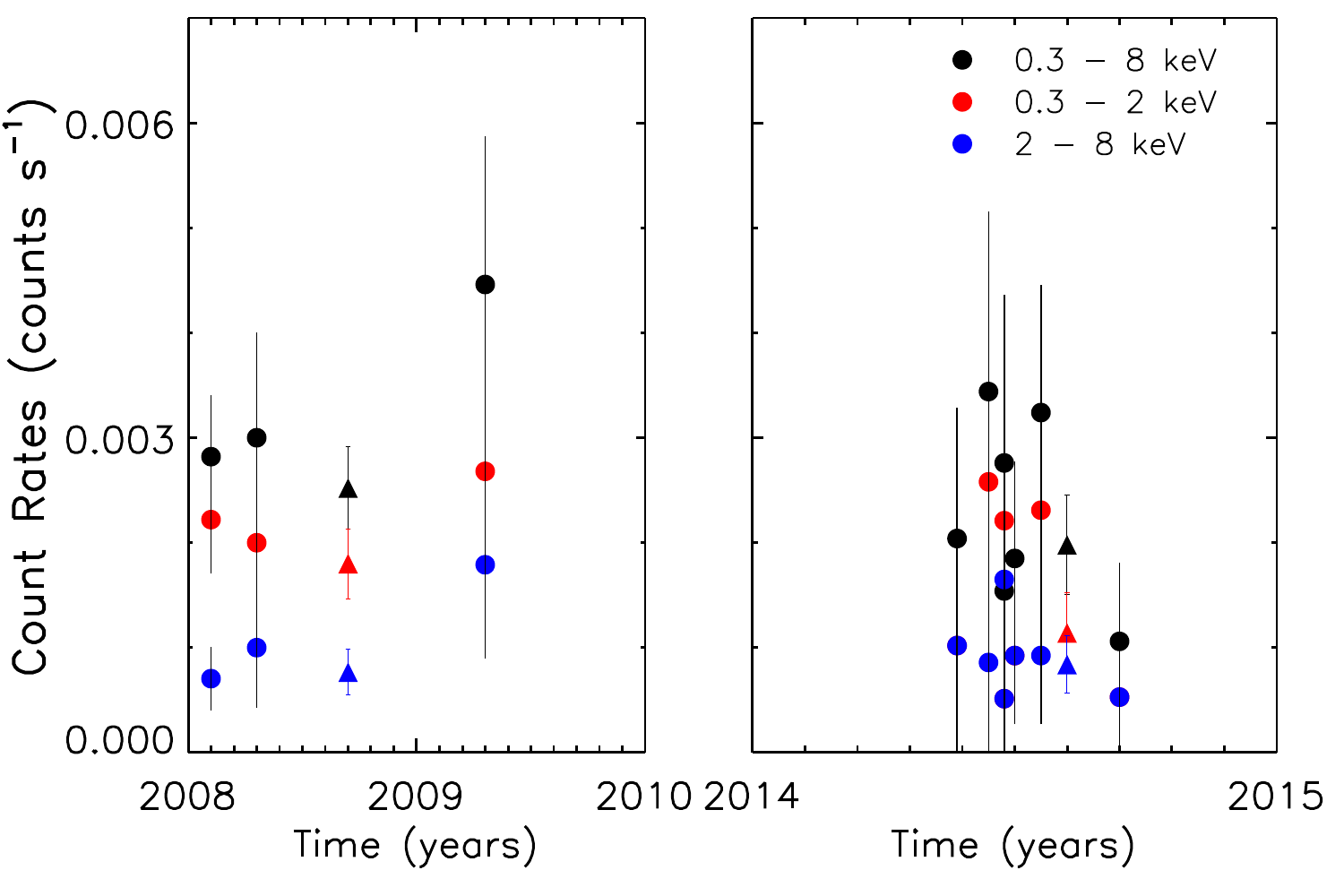}
}

{
(c)
\centering
\label{fig:sub:c}
\includegraphics[scale=0.35]{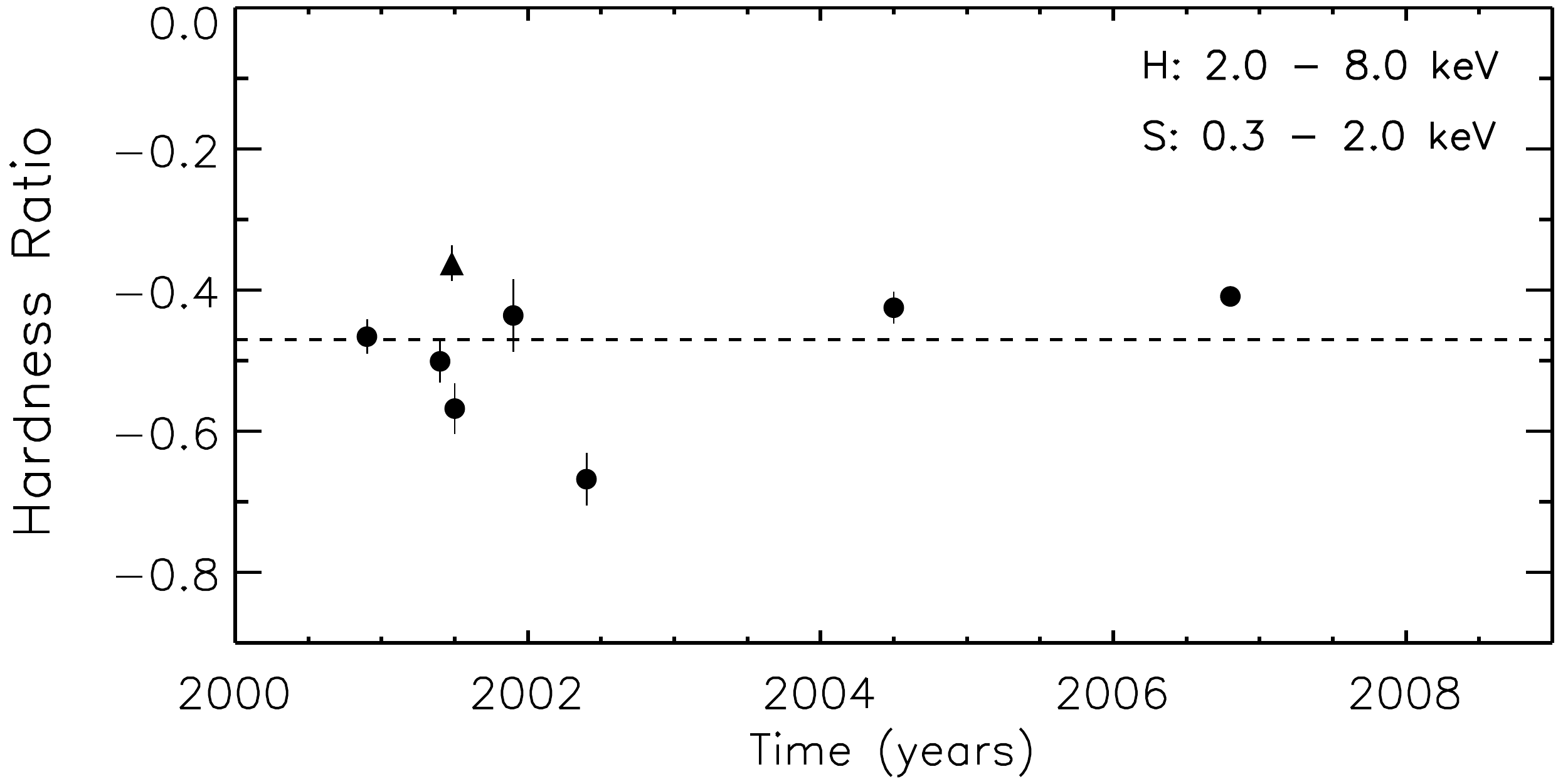}
\caption{The long-term light curve of NGC 4258 X-6 obtained using (a) {\it XMM-Newton} and {\it Chandra} and (b) {\it Swift} data. The counts were calculated in three energy ranges. In the upper left panel, the filled circles and the filled triangle represent {\it XMM-Newton} and {\it Chandra} data, respectively. In the upper right panel, the filled circles and filled triangles represent individual and combined  {\it Swift} data, respectively. (c) The long-term hardness evolution of NGC 4258 X-6 calculated with {\it XMM-Newton} (filled circles) and {\it Chandra} (filled triangle) data. The horizontal dashed line represents the average value of the hardness ratio calculated from the average count rate of the given energy bands.}
}
\end{figure*}

The light curve in Figure 3(a) shows that the count rate of the source in the total and soft bands changes by a factor of $\sim$2 and that in the hard band by a factor of $\sim$2.5 between 2001 and 2007. In Figure 3(b), since the individual {\it Swift} count rate has large error bars, the data sets taken in 2008$-$2009 (S1$-$S4) and 2014 (S5$-$S12) were combined and the count rates of X-6 were calculated as $\sim$0.0025 and 0.0020 counts s$^{-1}$ in the energy range 0.3$-$8 keV, respectively. The combined {\it Swift} count rates indicate a rather persistent behavior. The hard and soft count rates obtained with combined data sets are presented in Figure 3(b) as well.

The hardness ratios (HR), defined as $HR=(H-S)/(H+S)$, of X-6 were obtained. The long-term evolution of hardness ratio is given in Figure 3(c). As seen in the figure, X-6 has a noticeably softer HR value in observation XM5 but is nearly constant around the average ($\sim -$0.47) in the other observations. This is in agreement with the fact that the source has the lowest hard count rate in the XM5 data, which are given in Figure 3(a). 

The spectral fitting was performed using the {\it XMM-Newton} and {\it Chandra} data with {\scshape xspec} (version 12.8.1).\footnote{https://heasarc.gsfc.nasa.gov/xanadu/xspec/} All spectra were grouped to have a minimum of 20 counts per bin. In the {\it XMM-Newton} data, the EPIC pn and MOS spectra were fitted simultaneously by including a constant parameter to the fitted models. The constant values that were calculated from the XM7 data set were adopted and fixed while fitting the other {\it XMM-Newton} observations to achieve consistency. We fitted the 0.3$-$10 keV spectrum from each {\it XMM-Newton} and {\it Chandra} observation using absorbed PL and disk blackbody (DISKBB) models with two absorption components (using tbabs model in {\scshape xspec}). One of the absorption components was fixed to the Galactic value ($0.01\times10^{22}$ cm$^{-2}$; \citealt{dic90}), while the other one was set free to take into account the intrinsic absorption toward the source. The unabsorbed flux values were calculated in the energy range 0.3$-$10 keV using {\scshape cflux} in {\scshape xspec}. The spectral results of the fits for individual data sets are given in Table 3. The energy spectra of the ULX obtained using C1, XM6, and XM7 data are given in Figure 4 for PL and DISKBB models.

Considering the reduced $\chi^{2}$ values, both PL and DISKBB models provided statistically equivalent fits in XM4 and XM6 data. Also, XM3 data seem better fitted with DISKBB. But for the remaining data sets, the spectra of the source yielded relatively better fits with PL. The analyses revealed that X-6 generally has a hard spectrum ($\Gamma \sim$ 1.8$-$2.1). However, the spectrum becomes somewhat softer in XM5 data with a steeper photon index of $\Gamma \approx 2.4$ and a lower inner temperature of $T_{\mathrm{in}} \approx 0.8$ keV as derived from PL and DISKBB models, respectively. The calculated flux of X-6 between 0.3 and 10 keV is not constant throughout the observations. The luminosity of the source varies by a factor of 2 and it is in the range $L_{\mathrm{X}} \sim (1.1-2.2) \times 10^{39}$ erg s$^{-1}$ for PL and $L_{\mathrm{X}} \sim (0.5-1.2) \times 10^{39}$ erg s$^{-1}$ for DISKBB models. Furthermore, the absorbed PL+DISKBB composite model was fitted to the spectra. Statistically, the addition of a disk component did not improve the fit significantly.

The energy spectra of some ULXs may show curvature above 2 keV, which is expected from the optically thick corona \citep{sto06,gla09}. Therefore, to search for spectral curvature, we fitted the spectra of X-6 in XM6 and XM7 data with a broken PL model. The modeling gave the best-fit break energies as $\sim$2.9 and 3.6 keV, respectively. However, according to an F-test, the improvement of the fits over the PL model was $<2\sigma$.

A possible iron line was seen in the spectrum of the source in XM6 data. The observation of iron line emission in some ULX spectra could be interpreted to mean that the ULX is in its high state \citep{str07}. This line was fitted with an additional Gaussian line component to the PL model (see Figure 5). The best-fit model parameters for the Gaussian line were $E_{\mathrm{line}}=6.90_{-0.12}^{+0.28}$ and $\sigma = 0.16_{-0.15}^{+0.22}$. But the iron line is weak and the improvement of the fit was $<2\sigma$. This weak line was not detected in XM7 data, which have a longer exposure. We carried out two tests to check whether the line is associated with the ULX. For the first one, a combined spectrum was obtained by stacking all the pn data of all {\it XMM-Newton} data to increase the S/N around 7 keV. The line is not significantly detected in the combined spectrum. This is consistent with the possibility that the line emission appears only at certain epochs. For the second test, an image was obtained from pn data that was filtered between 6.6 and 7.5 keV to see whether those few line photons are centered on the position of the source or are spuriously contaminating the source region. The photons seem to center on the source. These results suggest that the iron line (if real) is a variable phenomenon associated with the ULX and not with diffuse emission from gas in the host galaxy.

Additionally, we tried to fit the spectra of X-6 with an absorbed DISKPBB model (also known as $p$-free disk or extended disk blackbody) to interpret the difference between standard disk and slim disk. Disk temperature has a radial dependence as $T$($R$) $\propto$ $R^{p}$, where $p$ is a free parameter. When $p = 0.75$ the standard disk model is obtained, and if $p < 0.75$ then radial advection becomes important. The spectral parameters for each observation are given in Table 4. The spectrum of X-6 is well modeled with a $p$-free model with a $p$ parameter that is indicative of a non-standard disk ($\sim$0.5), except for C1 data where the model indicated a $p$ parameter consistent with a standard disk. However, we note that the fit statistics are not good enough to distinguish between models. Additionally in XM1, XM2, XM4, and XM5 data the calculated inner disk temperatures ($T_{\mathrm{in}}$) were $\geq$ 3 keV. Therefore, we fixed the inner disk temperatures to the averaged value ($\sim$1.70 keV) while fitting these data (see Table 4).

We also calculated bolometric luminosity by integrating the DISKBB model fluxes between energies of 0.01 and 100 keV to obtain a plot of $L_\mathrm{{bol}}$--$T_{\mathrm{in}}$ (see Figure 6). The calculated luminosity values are given in column 9 of Table 4. The plot was fitted with a power-law relation and a $L_\mathrm{{bol}} \propto$ $T_{\mathrm{in}}^{1.5\pm{0.3}}$ was found with a correlation coefficient of $\sim$0.8, instead of $L_\mathrm{{bol}} \propto$ $T_{\mathrm{in}}^{4}$. This relation is expected from an optically thick standard accretion disk \citep{mak00}. Therefore, it seems rather difficult to interpret the emission of ULX X-6 as being the result of a standard disk.

We could not perform spectral fits to the {\it Swift} data, since the spectral quality was not good enough. However, the maximum 2$\sigma$ limit on the flux was calculated using the longest {\it Swift} data (S1) to be 3$\times10^{-13}$ erg cm$^{-2}$ s$^{-1}$ by fixing the parameters to the best PL parameters of XM7 data. This corresponds to a luminosity value of $L_{0.3-10} \approx 2 \times 10^{39}$ erg s$^{-1}$ at the adopted distance.

We examined the correlation between the best-fit model parameters (given in Table 3) and HR values in order to check the consistency of the fitting parameters. The plots are given in Figure 7. As seen in the upper two panels, the parameters $\Gamma$ and $T_{\mathrm{in}}$ parameters correlate well with HR. However, no correlation could be found between $N_{\mathrm{H}}$ values, obtained using PL and DISKBB models, and HR values.

\begin{figure*}
\centering
{
\label{fig:sub:a}
\includegraphics[scale=0.35, angle=270]{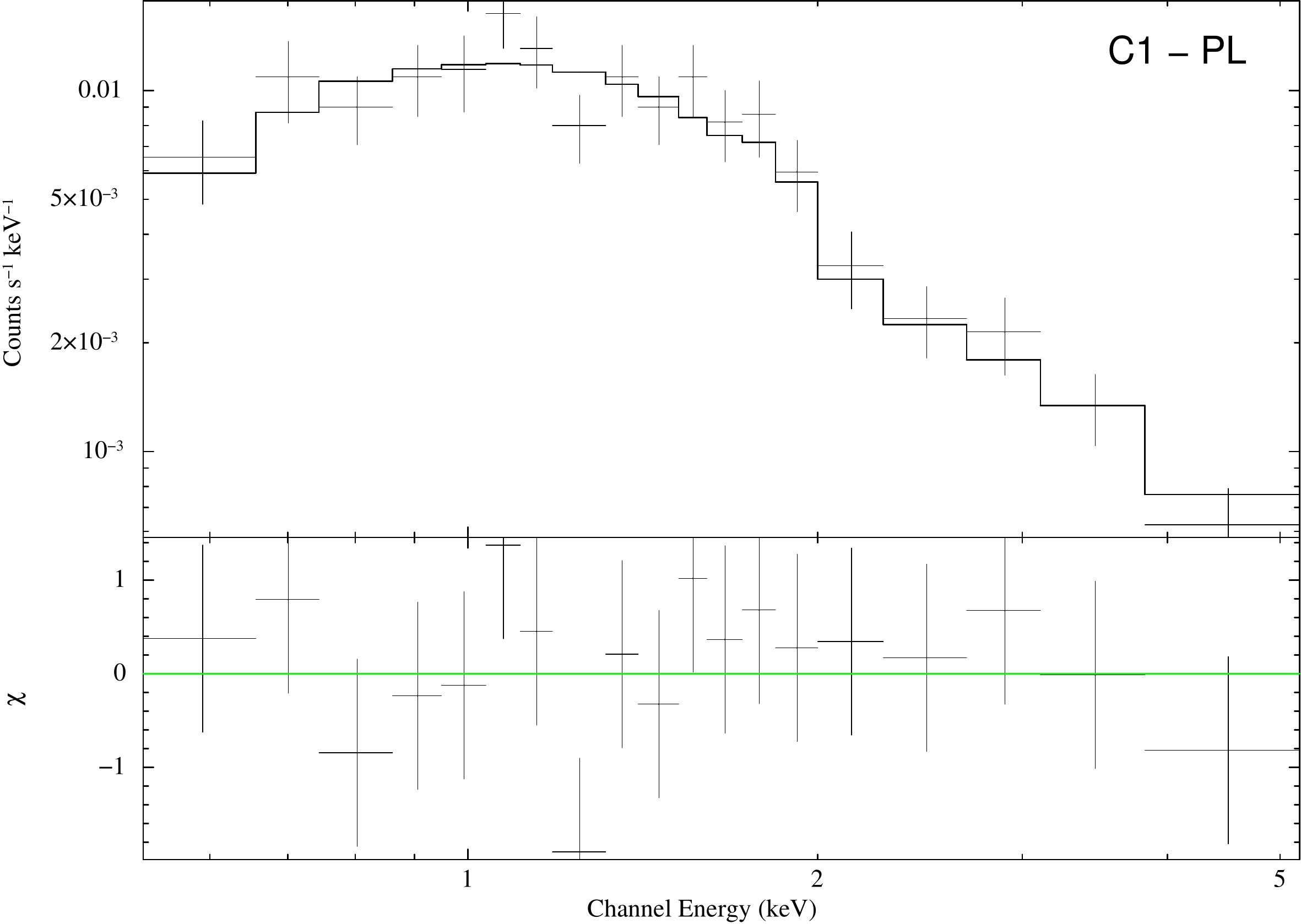}
}
{
\label{fig:sub:b}
\includegraphics[scale=0.35, angle=270]{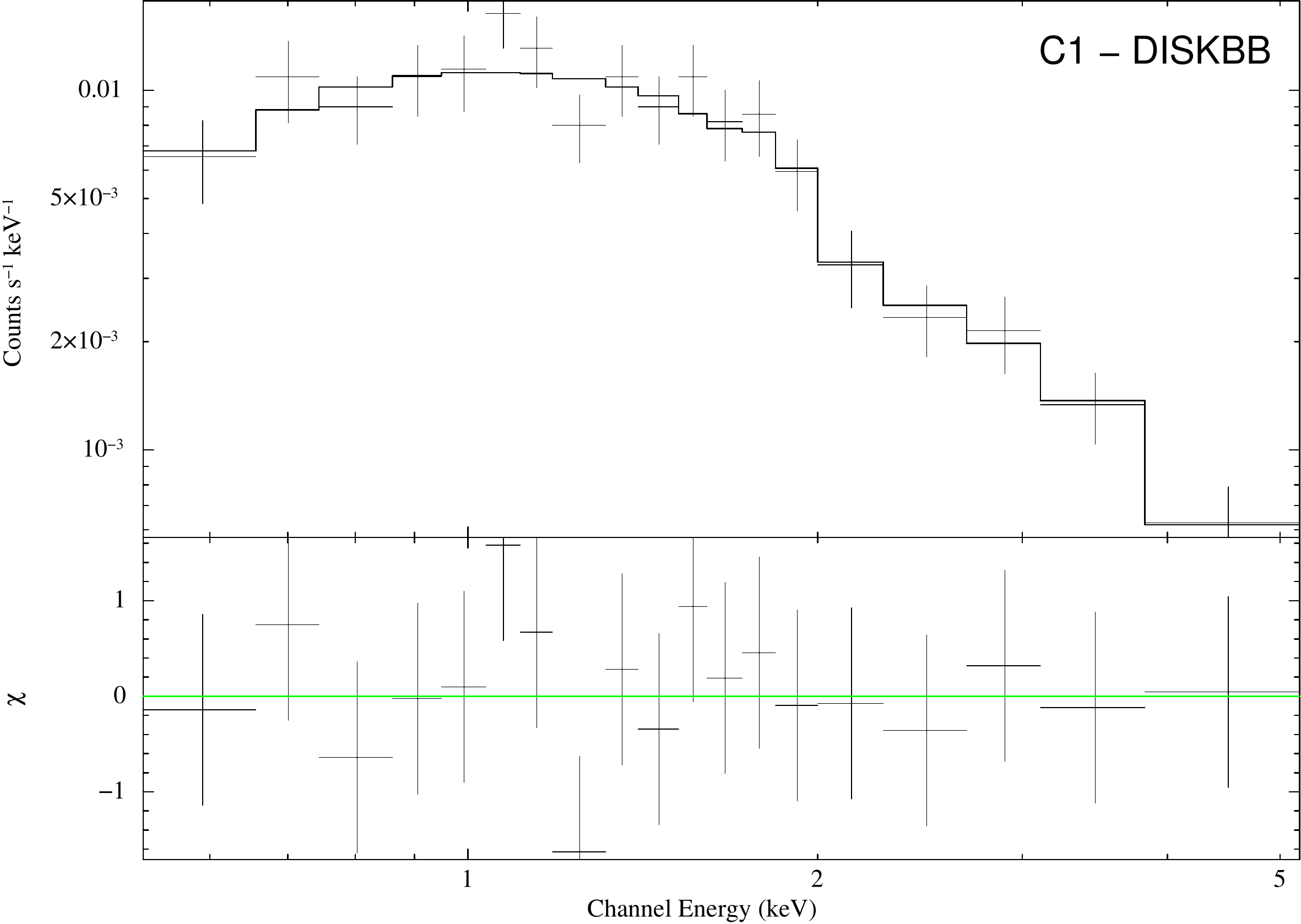}
}
{
\label{fig:sub:c}
\includegraphics[scale=0.35, angle=270]{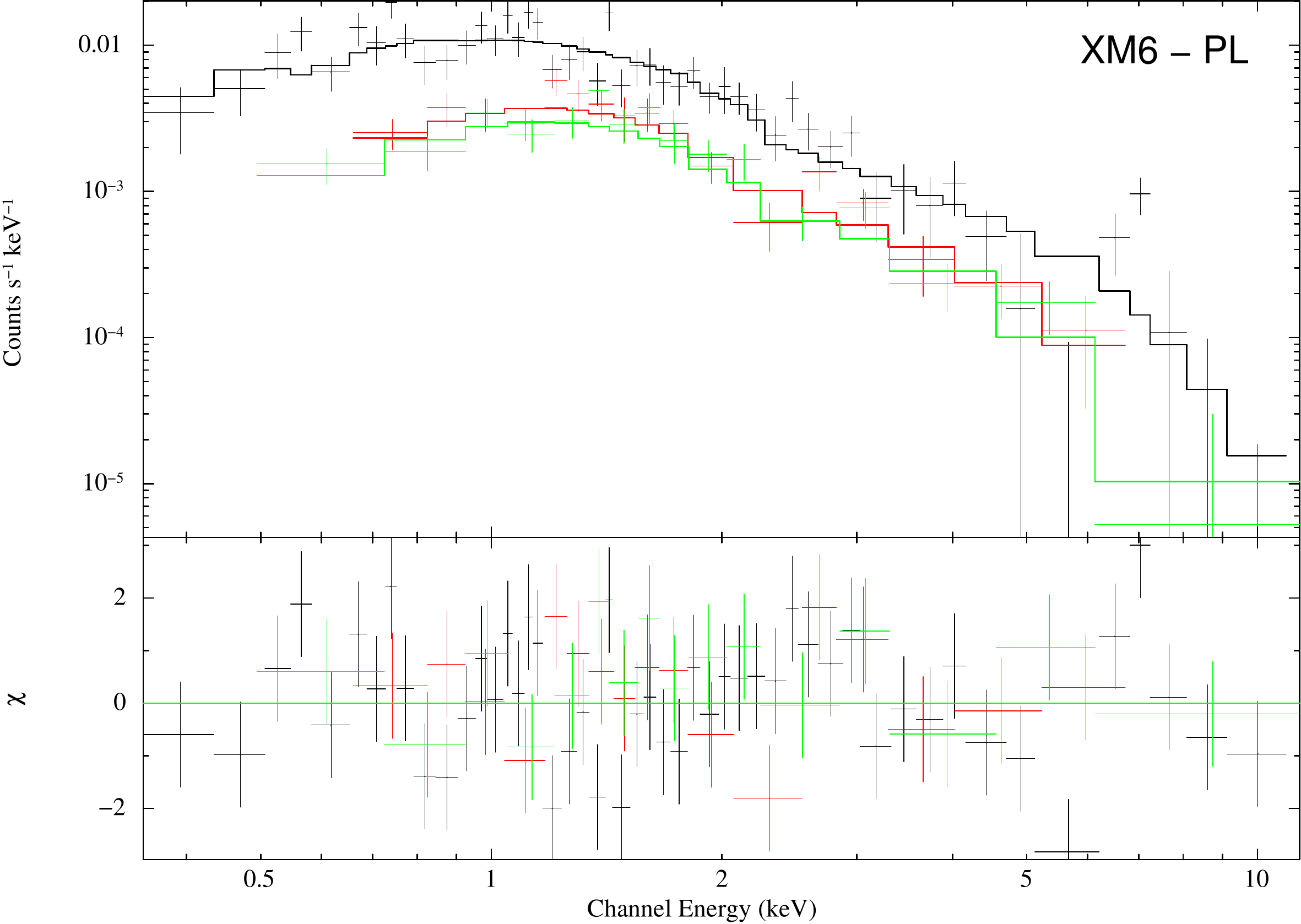}
}
{
\label{fig:sub:d}
\includegraphics[scale=0.35, angle=270]{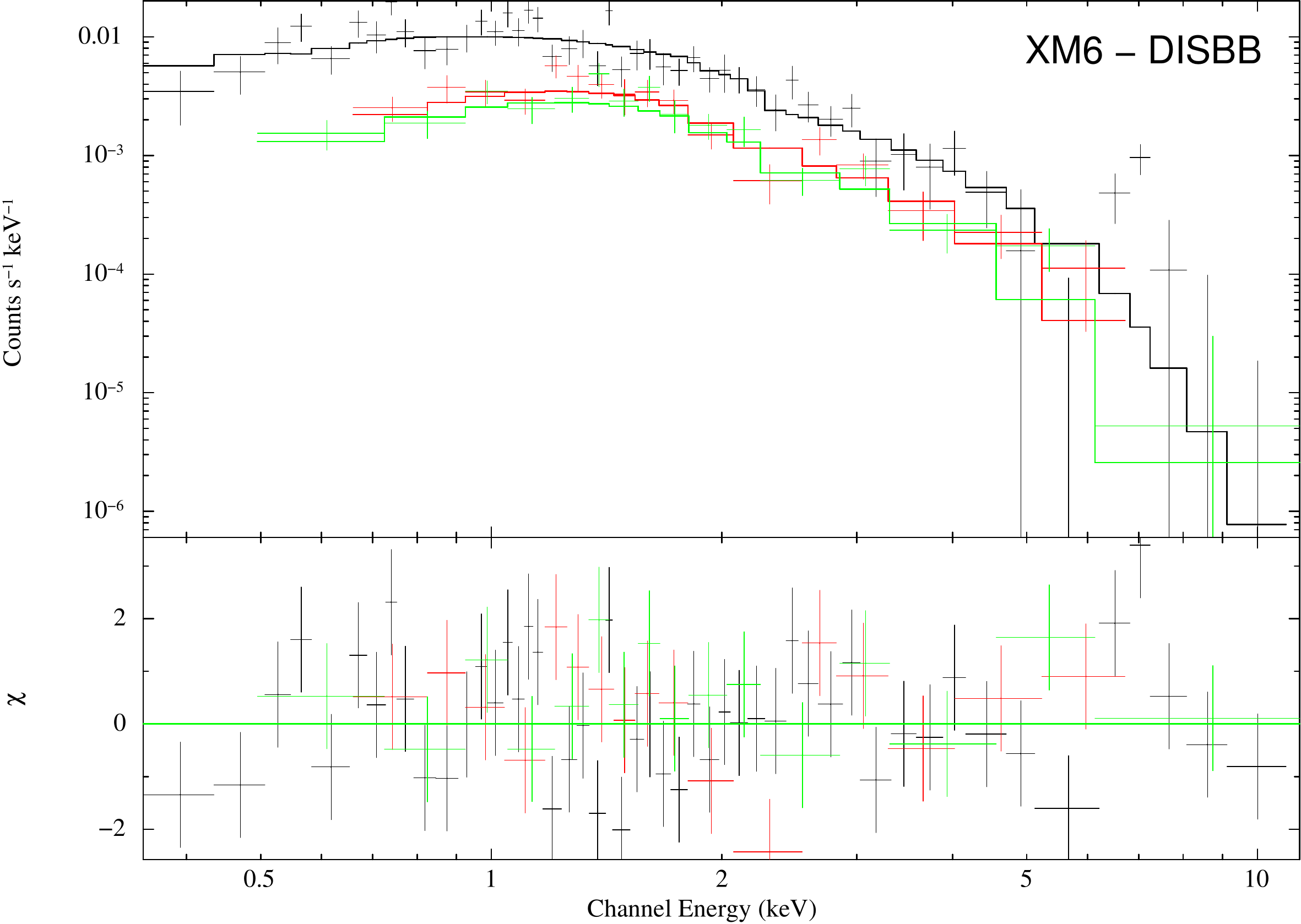}
}
{
\label{fig:sub:e}
\includegraphics[scale=0.35, angle=270]{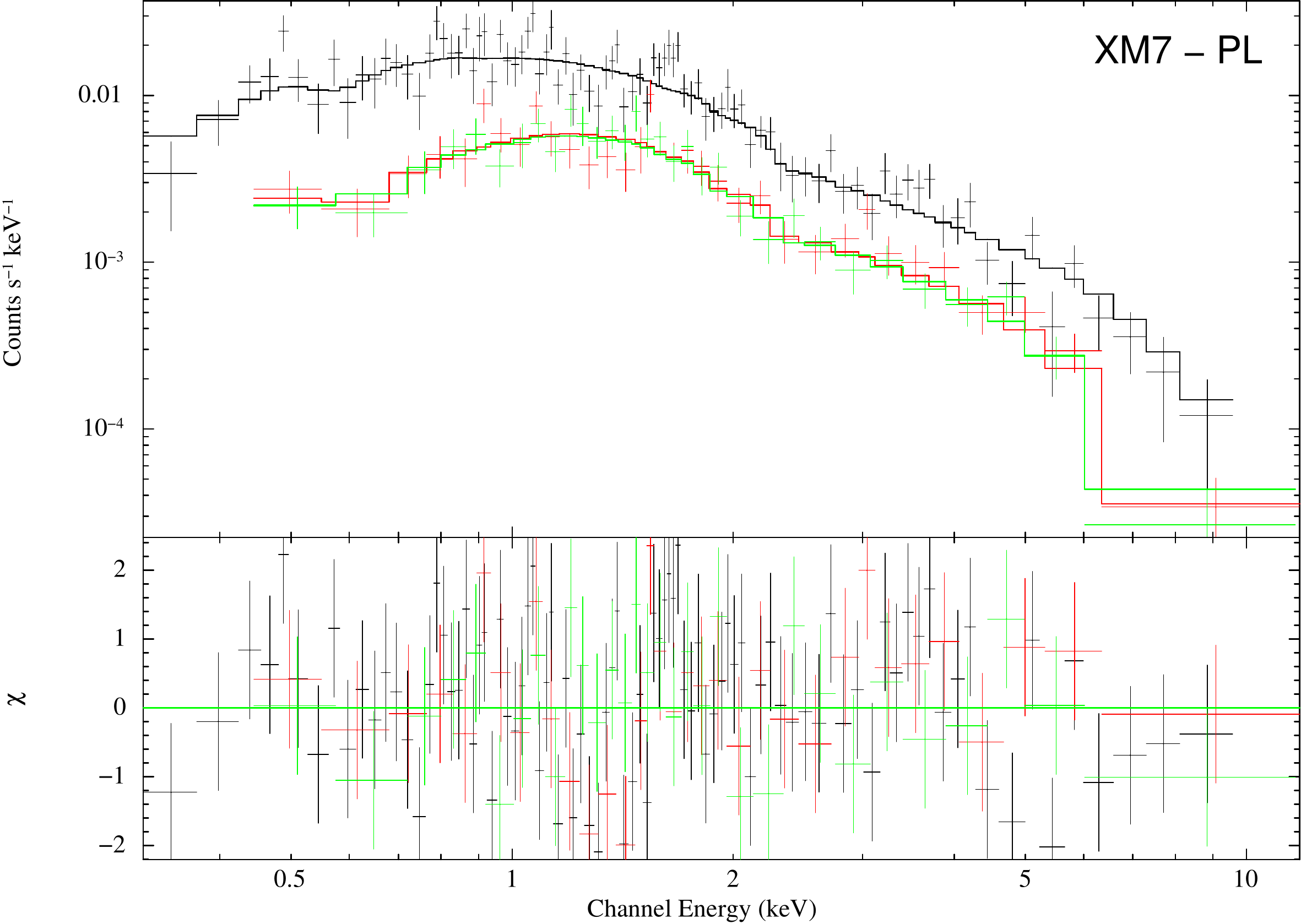}
}
{
\label{fig:sub:f}
\includegraphics[scale=0.35, angle=270]{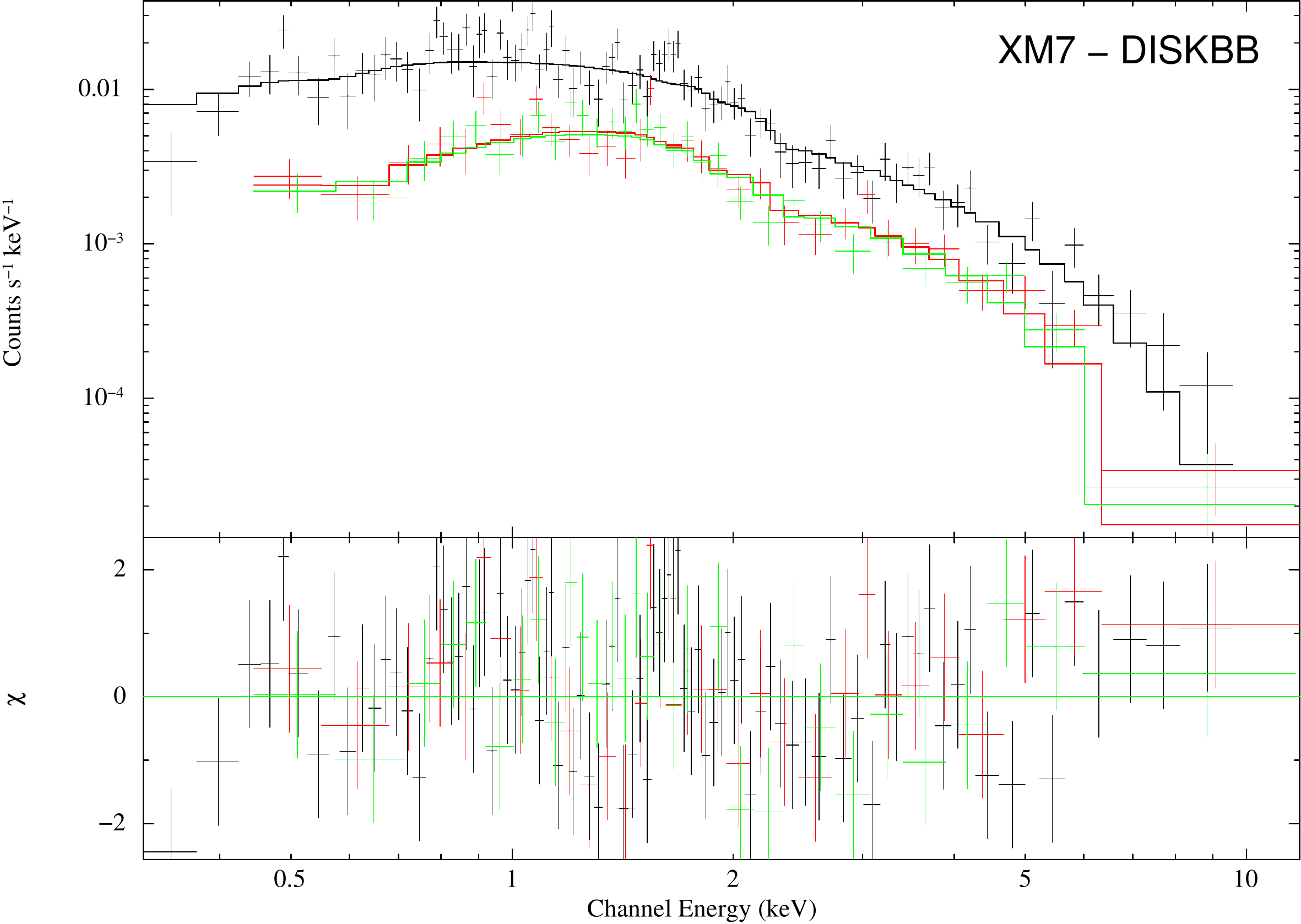}
\caption{Energy spectra of X-6 obtained using C1, XM6, and XM7 data. In {\it XMM-Newton} spectra, the black, green, and red data points represent EPIC pn, MOS1, and MOS2, respectively. The fitting models are denoted in the captions of each plot.}
}
\end{figure*}

\begin{figure}
\includegraphics[scale=0.35, angle=270]{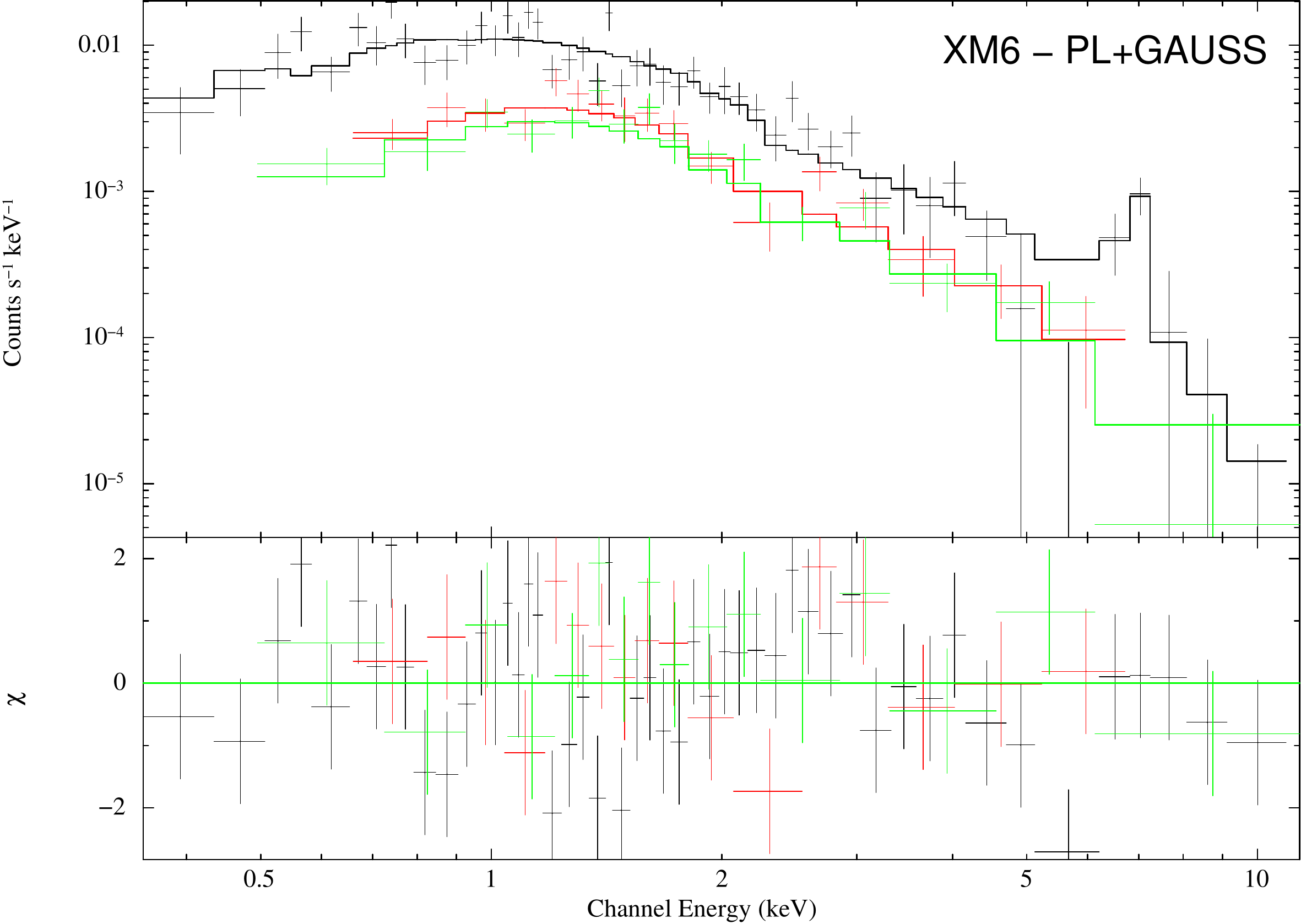}
\caption{The energy spectrum of X-6 in XM6 data. The spectrum was fitted with an absorbed PL+Gaussian model.}
\end{figure}

\begin{figure}
\includegraphics[width=\columnwidth]{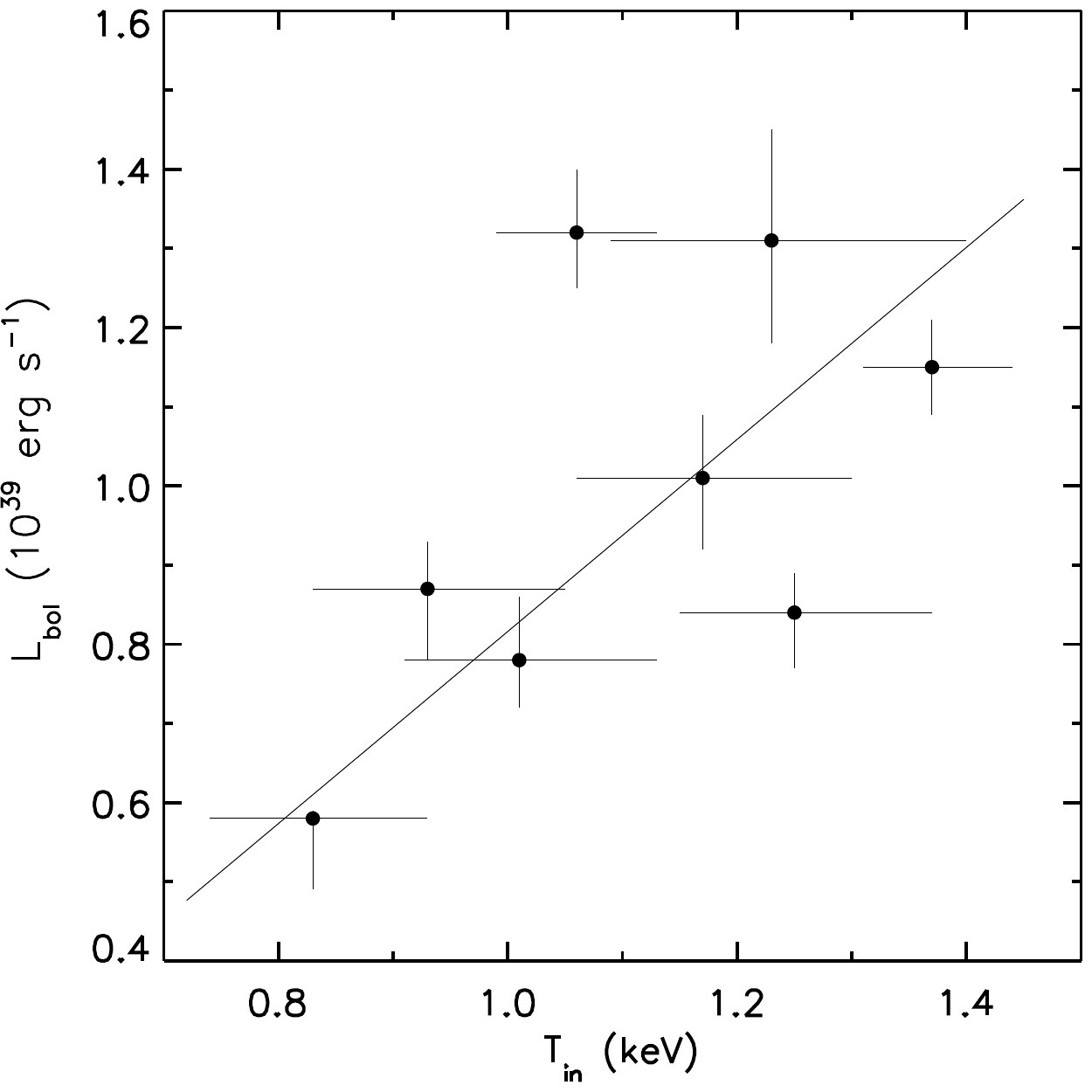}
\caption{$L_\mathrm{{bol}}$ derived from DISKBB models vs. $T_{\mathrm{in}}$. The black line represents the best-fit relation $L_\mathrm{{bol}} \propto$ $T_{\mathrm{in}}^{1.5\pm{0.3}}$ with a correlation coefficient of $\sim$0.8.}
\end{figure}

\begin{figure}
\includegraphics[width=\columnwidth]{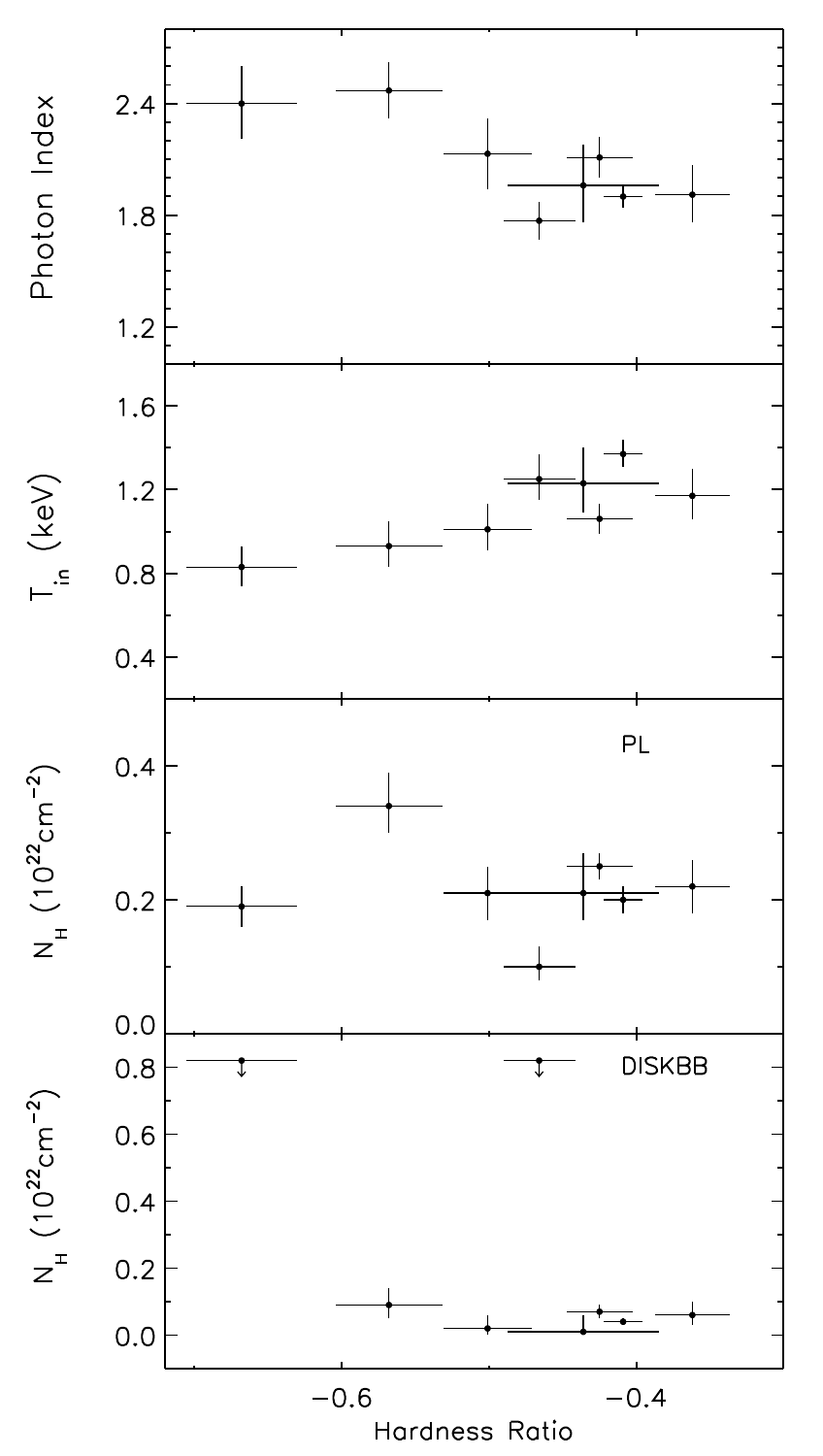}
\caption{The spectral parameters obtained with PL and DISKBB models vs. hardness ratio (HR).}
\end{figure}

\begin{deluxetable*}{c c c c c c c c c}
\tablewidth{0pt}
\tablecaption{X-Ray Spectral Fitting Parameters for X-6}
\tablehead{
\colhead{No.} & \colhead{$N_{\mathrm{H}}$} & \colhead{$\Gamma$} & \colhead{$T_{\mathrm{in}}$} & \colhead{$\chi^{2}$/dof} & \colhead{$N_{\mathrm{PL}}$\tablenotemark{a}} & \colhead{$N_{\mathrm{disk}}$\tablenotemark{b}} & \colhead{$L_{\mathrm{X}}$\tablenotemark{c}} &
\colhead{$L_{\mathrm{bol}}$\tablenotemark{d}} \\
 & $(10^{22}$ cm$^{-2}$) & & (keV) & & (10$^{-5}$) & (10$^{-3}$) & (10$^{39}$ erg s$^{-1}$) & (10$^{39}$ erg s$^{-1}$)  }
\startdata
\multicolumn{9}{c}{tbabs*powerlaw}\\
\tableline
XM1 & $0.10_{-0.02}^{+0.03}$ & $1.77_{-0.10}^{+0.10}$ & - & 40.11/33 & $2.59_{-0.18}^{+0.18}$ & - & $1.21_{-0.09}^{+0.08}$ & -  \\ 
XM2 & $0.21_{-0.04}^{+0.04}$ & $2.13_{-0.19}^{+0.19}$ & - & 16.76/17 & $3.66_{-0.32}^{+0.32}$ & - & $1.37_{-0.11}^{+0.12}$ & -  \\
C1 & $0.22_{-0.04}^{+0.04}$ & $1.91_{-0.15}^{+0.16}$ & - &  10.39/15 & $3.74_{-0.33}^{+0.33}$ & - & $1.57_{-0.14}^{+0.14}$ & -  \\   
XM3 & $0.34_{-0.04}^{+0.05}$ & $2.47_{-0.15}^{+0.15}$ & - & 15.67/22 & $5.49_{-0.48}^{+0.48}$ & - & $1.80_{-0.15}^{+0.17}$ & -  \\
XM4 & $0.21_{-0.04}^{+0.06}$ & $1.96_{-0.20}^{+0.22}$ & - & 35.13/28 & $5.05_{-0.53}^{+0.53}$ & - &  $2.06_{-0.21}^{+0.21}$ & -   \\
XM5 & $0.19_{-0.03}^{+0.03}$ & $2.40_{-0.19}^{+0.20}$ & - & 30.49/28 & $3.32_{-0.29}^{+0.29}$  & - & $1.11_{-0.10}^{+0.10}$ & -  \\
XM6 & $0.25_{-0.02}^{+0.02}$ & $2.11_{-0.11}^{+0.11}$ & - & 101.26/79 &  $5.84_{-0.32}^{+0.32}$ & - & $2.20_{-0.12}^{+0.12}$  & - \\
XM7 & $0.20_{-0.02}^{+0.02}$ & $1.90_{-0.06}^{+0.06}$ & - &  160.34/150 & $3.87_{-0.19}^{+0.19}$ & - & $1.65_{-0.08}^{+0.07}$ & -  \\
\cutinhead{tbabs*diskbb}
XM1 & $<$0.82  & - & $1.25_{-0.10}^{+0.12}$  & 51.67/33 & - & $2.21_{-0.16}^{+0.16}$  &  $0.80_{-0.06}^{+0.05}$ & $0.84_{-0.07}^{+0.05}$ \\
XM2 & $0.02_{-0.02}^{+0.04}$ & - & $1.01_{-0.10}^{+0.12}$  & 19.07/17 & - & $4.96_{-0.43}^{+0.43}$ & $0.74_{-0.07}^{+0.06}$ & $0.78_{-0.06}^{+0.08}$  \\
C1 & $0.06_{-0.03}^{+0.04}$  & - & $1.17_{-0.11}^{+0.13}$  &  8.18/15 & - & $3.48_{-0.30}^{+0.30}$  & $0.96_{-0.09}^{+0.08}$ & $1.01_{-0.09}^{+0.08}$ \\ 
XM3 & $0.09_{-0.04}^{+0.05}$ & - & $0.93_{-0.10}^{+0.12}$  & 20.26/22 & - & $7.50_{-0.66}^{+0.66}$ & $0.82_{-0.07}^{+0.07}$  & $0.87_{-0.09}^{+0.06}$ \\
XM4 & $0.01_{-0.01}^{+0.05}$ & - & $1.23_{-0.14}^{+0.17}$  & 35.97/28 & - & $3.74_{-0.39}^{+0.39}$ & $1.23_{-0.12}^{+0.14}$ & $1.31_{-0.13}^{+0.14}$ \\
XM5 & $<$0.82 & - & $0.83_{-0.09}^{+0.10}$  & 35.18/28 & - & $7.33_{-0.66}^{+0.66}$ & $0.53_{-0.05}^{+0.05}$ & $0.58_{-0.09}^{+0.01}$ \\
XM6 & $0.07_{-0.02}^{+0.02}$ & - & $1.06_{-0.07}^{+0.07}$  & 101.28/79 & - & $6.87_{-0.37}^{+0.37}$ & $1.24_{-0.06}^{+0.06}$ & $1.32_{-0.07}^{+0.08}$ \\
XM7 & $0.04_{-0.01}^{+0.01}$ & - & $1.37_{-0.06}^{+0.07}$  & 174.51/150 & - & $2.12_{-0.10}^{+0.10}$ & $1.09_{-0.05}^{+0.06}$ & $1.15_{-0.06}^{+0.06}$ \\
\enddata
\tablenotetext{a}{Normalization parameter of the PL model in units of photon cm$^{-2}$ s$^{-1}$ keV$^{-1}$ at 1 keV.}
\tablenotetext{b}{Normalization parameter of the DISKBB model. N$_{\mathrm{disk}}=[(r_{\mathrm{in}}/\mathrm{km})/(D/10 \mathrm{kpc})]^{2} \times \cos i$, where $r_{\mathrm{in}}$ is the apparent inner disk radius, $D$ is the distance to the source, and $i$ is the inclination of the disk.}
\tablenotetext{c}{The luminosity values were calculated using a distance of 7.7 Mpc \citep{swa11}.}
\tablenotetext{d}{The bolometric luminosity values were calculated in the energy range 0.01$-$100 keV.}
\end{deluxetable*}

\begin{deluxetable*}{c c c c c c c }
\tablewidth{0pt}
\tablecaption{X-Ray Spectral Parameters Obtained with the DISKPBB Model}
\tablehead{
\colhead{No.} & \colhead{$N_{\mathrm{H}}$} & \colhead{$T_{\mathrm{in}}$} & $p$ & \colhead{$\chi^{2}$/dof} & \colhead{$N_{\mathrm{disk}}$\tablenotemark{a}} & \colhead{$L_{\mathrm{X}}$\tablenotemark{b}} \\
 & $(10^{22}$ cm$^{-2}$) & (keV) & & & (10$^{-4}$) & (10$^{39}$ erg s$^{-1}$)}
\startdata
\multicolumn{7}{ c }{tbabs*diskpbb} \\
\tableline
XM1 & $0.05_{-0.02}^{+0.03}$ & 1.70 & $0.60_{-0.03}^{+0.03}$ & 44.17/33 & $3.49_{-0.25}^{+0.25}$ & $0.92_{-0.07}^{+0.06}$  \\ 
XM2 & $0.16_{-0.03}^{+0.04}$ & 1.70 & $0.52_{-0.03}^{+0.03}$ & 17.04/17 & $1.93_{-0.16}^{+0.16}$ & $1.11_{-0.10}^{+0.10}$  \\
C1 & $0.05_{-0.03}^{+0.04}$ & $1.16_{-0.11}^{+0.13}$ & $0.76_{-0.07}^{+0.10}$ &  8.18/14 & $37.39_{-0.33}^{+0.33}$ & $0.95_{-0.08}^{+0.09}$  \\   
XM3 & $0.27_{-0.04}^{+0.05}$ & $1.52_{-0.39}^{+0.78}$ & $0.50_{-0.02}^{+0.02}$ & 16.30/21 & $2.88_{-0.25}^{+0.25}$ & $1.32_{-0.11}^{+0.12}$  \\
XM4 & $0.10_{-0.04}^{+0.05}$ & 1.70 & $0.59_{-0.04}^{+0.05}$ & 35.34/28 & $5.29_{-0.55}^{+0.55}$ & $1.48_{-0.14}^{+0.18}$   \\
XM5 & $0.16_{-0.03}^{+0.03}$ & 1.70 & $0.48_{-0.02}^{+0.03}$ & 30.86/28 & $0.93_{-0.08}^{+0.08}$ & $0.94_{-0.08}^{+0.09}$   \\
XM6 & $0.18_{-0.02}^{+0.02}$ & $1.54_{-0.19}^{+0.24}$ & $0.55_{-0.02}^{+0.02}$ & 99.06/78 &  $5.96_{-0.32}^{+0.32}$ & $1.66_{-0.09}^{+0.09}$  \\
XM7 & $0.16_{-0.02}^{+0.02}$ & $2.58_{-0.34}^{+0.46}$ & $0.55_{-0.01}^{+0.01}$ &  156.23/149 & $0.66_{-0.03}^{+0.03}$ & $1.45_{-0.08}^{+0.07}$ \\
\enddata
\tablenotetext{a}{Normalization parameter of the DISKPBB model. N$_{\mathrm{disk}}=[(r_{\mathrm{in}}/\mathrm{km})/(D/10 \mathrm{kpc})]^{2} \times \cos i$, where $r_{\mathrm{in}}$ is the apparent inner disk radius, $D$ is the distance to the source, and $i$ is the inclination of the disk.}
\tablenotetext{b}{The luminosity values were calculated using a distance of 7.7 Mpc \citep{swa11}.}
\end{deluxetable*}

\subsection{{\it HST}}

We have analyzed {\it HST}/ACS/WFC archival data listed in Table 2 to investigate the optical counterpart of X-6. The position of the ULX on the {\it HST}/ACS/WFC images was derived as a result of relative astrometric correction (see Section 2.2). There is one relatively bright extended object and another faint object within the error circle. This bright extended object could also be two stars. By carefully examining the images of F435W and F814W, we considered this possibility and label the three sources within the error circle of X-6 as "source 1," "source 2," and "source 3" (see Figure 2). Since the region is crowded, point-spread function (PSF) photometry was performed instead of aperture photometry. Three distinct sources were detected by PSF photometry. Hence, we have analyzed these three sources as possible optical counterparts of X-6. 

The PSF photometry was performed with the {\scshape dolphot} software version 2.0 \citep{dol00} using the {\it HST}/ACS/WFC module. The FITS files (*flt.fits and *.drz.fits) were retrieved from the {\it HST} data archive.\footnote{https://archive.stsci.edu/hst/search.php} Standard image reduction algorithms (bias and dark current subtraction, flat fielding) have been applied to the observations. The {\scshape acsmask} and {\scshape splitgroups} tasks were used to mask out all the bad pixels and split the multi-image FITS files into a single file per chip, respectively. Then the {\scshape dolphot} task was used for source detection, photometry, and photometric conversion. The {\scshape dolphot} task gives standard magnitudes using the conversion method as described by \citet{sir05}. This task was used for photometry on the images by taking the F435W drizzled image as the positional reference. The Galactic extinction along the direction to NGC 4258 is $E(\bv)=0.016$ mag \citep{sch98}. \citet{mac06} derived the extragalactic extinction for NGC 4258 in the range of $0.05 \leq E(\bv) \leq 0.28$ mag using 69 Cepheids. The mean extinction obtained from a few Cepheids close to the region around the ULX is $E(\bv)=0.05$ mag. This value is assumed to be more appropriate for the calculation of extinction-corrected magnitudes. Both $E(\bv)=0.016$ mag and $E(\bv)=0.05$ mag yielded compatible results. For this reason, we excluded the extinction effect of the galaxy and adopted the Galactic extinction for the reddening correction. The reddening-corrected instrumental VEGA magnitudes, Johnson magnitudes, colors, and absolute magnitudes are listed in the Table 5. 

On the other hand, we also analyzed the {\it HST}/ACS/WFC F606W archive images to check the optical variability of the counterpart candidates. These observations were performed on 2005 March 7, 9, and 10. The candidates that were found in the other {\it HST} filters were clearly identified in each observation. The VEGA magnitudes were calculated as 23.592$\pm$0.047, 23.581$\pm$0.042, 23.126$\pm$0.032 for source 1, 22.564$\pm$0.020, 22.659$\pm$0.021, 22.616$\pm$0.020 for source 2, and 23.001$\pm$0.031, 23.069$\pm$0.030, 23.126$\pm$0.032 for source 3. Source 1 shows significant variability ($\Delta m_{\mathrm{F606W}} =$ 0.481$\pm$0.056), while source 2 and 3 do not show notable variability ($\Delta m_{\mathrm{F606W}}$ $<$ 0.1 mag).

\begin{deluxetable}{c c c c  }
\tablewidth{0pt}
\tablecaption{Magnitude values of the Three Possible ULX Counterparts Obtained with the {\it HST}/ACS/WFC data}
\tablehead{
\colhead{Source No.} & \colhead{Filter} & \colhead{VEGA mag.} & \colhead{Johnson Mag.}}
\startdata
Source 1 &  F435W ($B$) & $24.146\pm 0.046$ & $24.198\pm 0.046$  \\
& F555W ($V$) & $24.167\pm 0.048$ & $24.139\pm 0.048$  \\
& F814W ($I$) & $24.364\pm 0.073$ & $24.349\pm 0.073$  \\
& $(B-V)_{0}$ & & $0.059\pm 0.066$  \\
& $(V-I)_{0}$ & & $-0.210\pm 0.087$   \\
& $M_{V}$ & & $-5.32\pm 0.048$   \\
Source 2 & F435W ($B$) & $22.479\pm 0.017$ & $22.528\pm 0.017$ \\
&F555W ($V$) & $22.482\pm 0.017$ & $22.450\pm 0.017$  \\
& F814W ($I$) & $22.508\pm 0.022$ & $22.494\pm 0.022$  \\
& $(B-V)_{0}$ & & $0.078\pm 0.024$  \\
& $(V-I)_{0}$ & & $0.044\pm 0.028$   \\
& $M_{V}$ & & $-7.01\pm 0.017$   \\
Source 3 & F435W ($B$) & $23.682\pm 0.034$ & $23.732\pm 0.034$ \\
&F555W ($V$) & $23.644\pm 0.034$ & $23.617\pm 0.034$  \\
& F814W ($I$) & $23.969\pm 0.054$ & $23.953\pm 0.054$ \\
& $(B-V)_{0}$ & & $0.115\pm 0.048$  \\
& $(V-I)_{0}$ & & $-0.336\pm 0.064$   \\
& $M_{V}$ & & $-5.84\pm 0.034$  
\enddata
\tablecomments{The extinction-corrected magnitudes were derived in the {\it HST}/ACS/WFC VEGA magnitude system and in the Johnson--Cousins (UBVRI) system.}
\end{deluxetable}

We obtained two color-magnitude diagrams (CMDs) as F555W versus F435W -- F555W and F814W versus F555W -- F814W to estimate the age of the sources and the cluster (See Figure 8). For the CMDs, the stars within 5$\arcsec$ radius with S/N $>$ 4 were selected. The PARSEC isochrones\footnote{http://stev.oapd.inaf.it/cgi-bin/cmd} of \citet{bre12} were used in the CMDs, based on the updated version of the code used to compute stellar tracks. The metallicity of NGC 4258 has been adopted from \citet{kud13} as Z $=0.011$ to obtain the isochrones. The Galactic reddening $E(\bv)$ of this region is given from dust maps as 0.016 according to \cite{sch98}. A distance modulus of 29.4 mag (using the  distance of 7.7 Mpc) was used to plot the CMDs. The PARSEC isochrones corresponding to the Z value have been overplotted in Figure 8. The selected nearby bright stars within the 5$\arcsec$ region have almost the same reddening values, which show that they could be in the same cluster. On the other hand, field stars have different color indices and reddening values. According to the CMDs, we are able to determine the age of the sources and the cluster as $<$50 MYr, comparable to the other clusters around ULXs (e.g. \citealt{abo07,gri11})

\begin{figure*}
(a)
{
\label{fig:sub:a}
\includegraphics[scale=0.65]{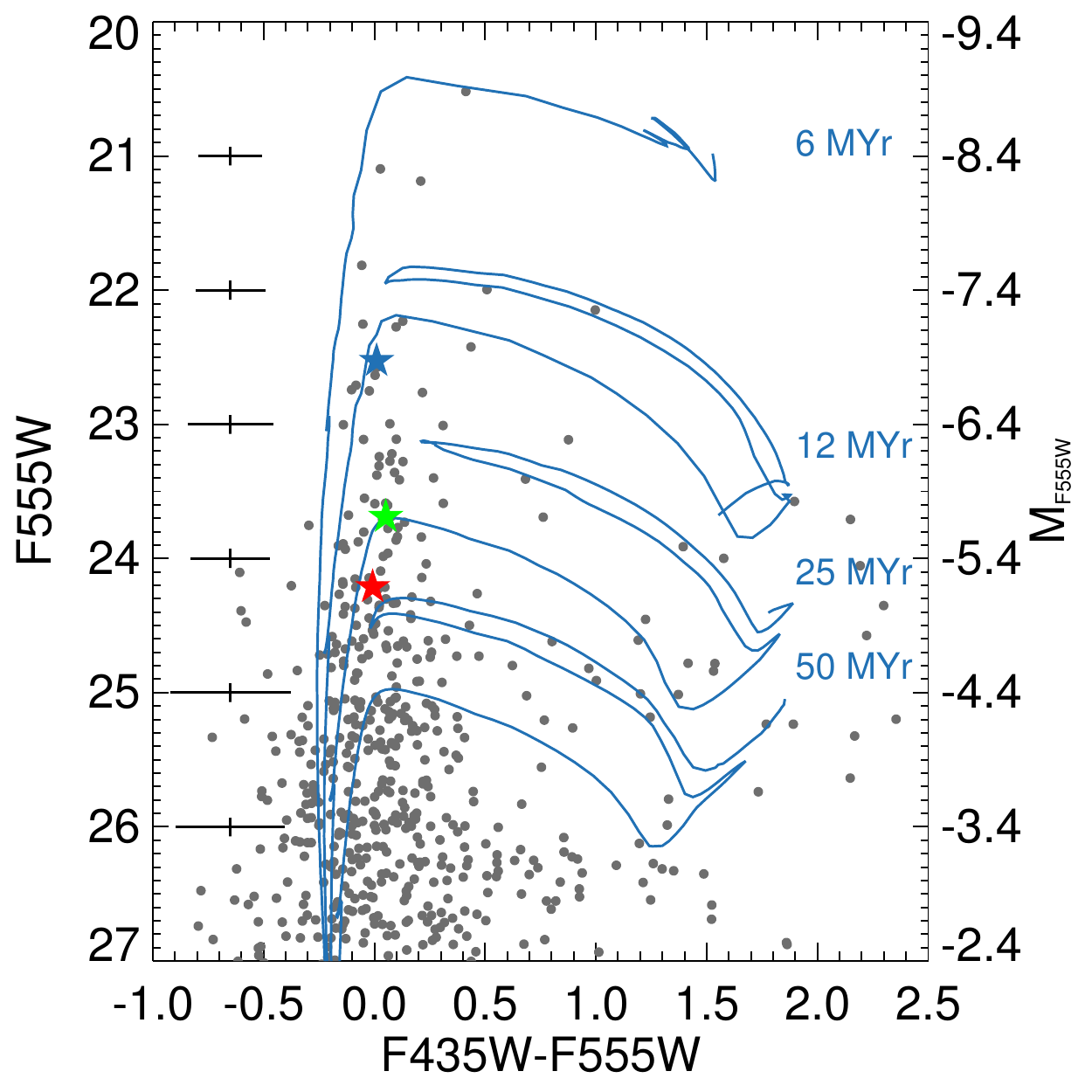}
}
(b)
{
\label{fig:sub:b}
\includegraphics[scale=0.65]{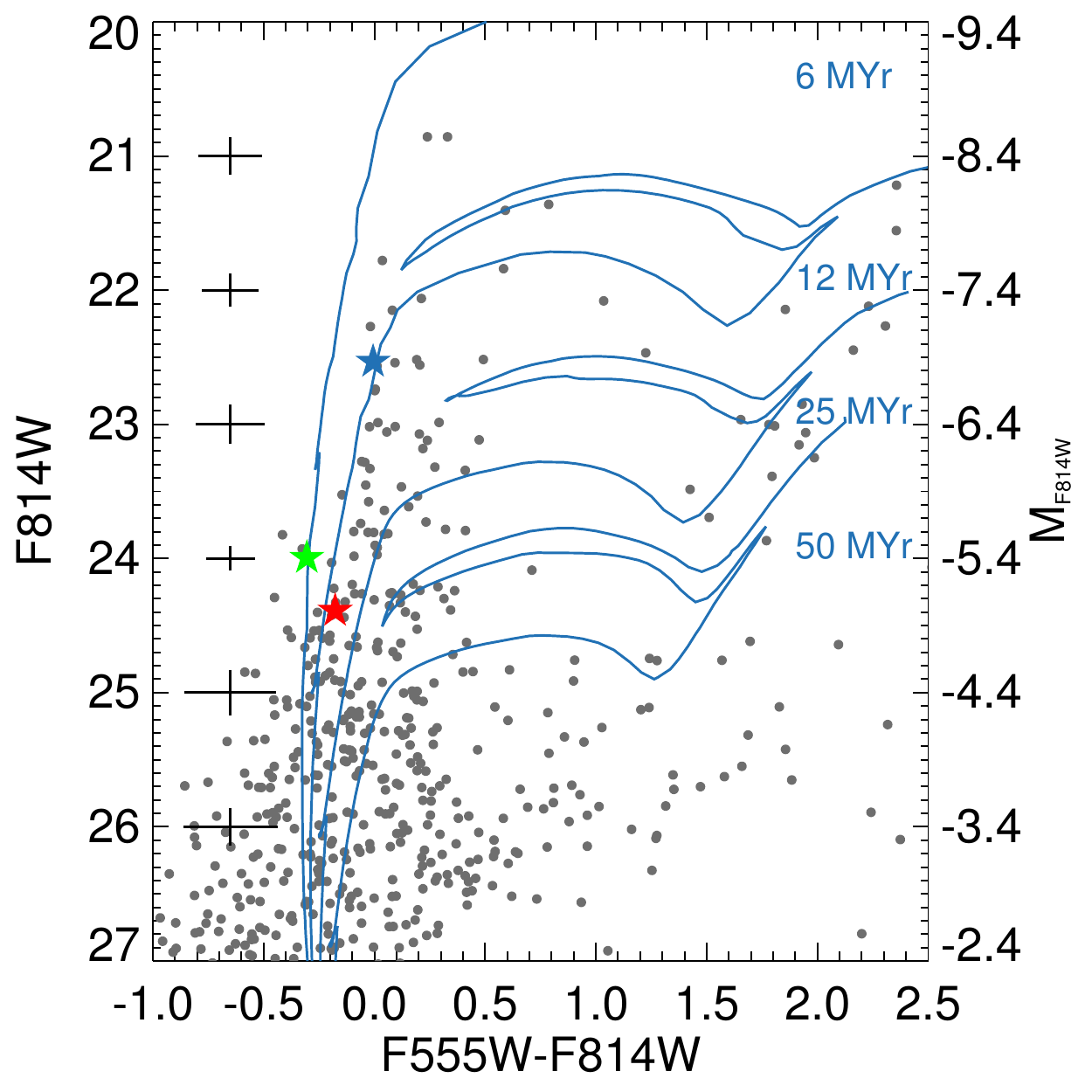}
\caption{{\it HST}/ACS/WFC CMDs for the stars in a region of radius 5$\arcsec$ around the ULX counterpart candidates. PARSEC isochrones for different ages and the mean magnitude errors are overplotted. All magnitudes have been corrected for extinction of $E(\bv)=0.016$ mag. The red, blue, and green stars represent sources 1--3, respectively.}
}
\end{figure*}

\section{Discussion and Summary}

We examined the X-ray temporal and spectral properties of X-6 using seven {\it XMM-Newton}, one {\it Chandra}, and 12 {\it Swift} observations available in the archives. Also, the optical counterpart of X-6 was investigated using the archival {\it HST}/ACS/WFC data. With the help of the simultaneous multi-band {\it HST}/ACS/WFC data, the CMDs for optical counterpart candidates and the cluster members have been obtained. 

As seen in Figure 3(a) and 3(c), X-6 exhibits possible spectral variations. However, the light curves from the {\it Swift} data do not show significant variation and indicate a rather persistent behaviour (see Figure 3(b)). Most notably, X-6 has the lowest HR value in XM5 data. The source has a steeper PL photon index ($\Gamma \sim 2.40$) and the lowest inner disk temperature ($T_{\mathrm{in}} \sim 0.83$ keV) at this epoch. If we assume that the source emits at the Eddington limit in XM6 data (in which the spectrum gives the highest $L_{\mathrm{X}} = 2.2 \times 10^{39}$ erg s$^{-1}$) the mass of the compact object in this system is found to be $M_{\mathrm{BH}}\sim 15 M_{\sun}$. The luminosity of X-6 changes by a factor $\sim$2 during these variations. Nonetheless, this variation in luminosity seems not to correlate with the canonical BHB states. Generally in Galactic BHBs, the luminosities are usually lower in the hard state, higher in the thermal (soft) state, and switch to a very high value in the steep PL state. However, there are some Galactic BHBs and ULXs that do not show similar behavior (e.g. XTE J1550-564, \citealt{rem06}; NGC 1313 X-2, \citealt{fen06}; IC 342 X-1, \citealt{mar14}; NGC 4736 X-2, \citealt{avd14}).

Taking into account that the ULXs are different than usual Galactic BHBs, each might harbour a stellar-mass BH with a supercritical accretion disk (SCAD) \citep{fab15} similar to GRS 1915+105 \citep{vie10}. In SCAD scenario the disk is expected to be slim ($H/R \sim 1$) within a spherization radius ($r_{\mathrm{sp}}$) and the temperature is expected to depend on the radius as $R^{1/2}$ \citep{pou07}. We tried to fit the X-ray spectra of X-6 with a $p$-free model in {\scshape xspec} to determine whether X-6 has slim disk properties \citep{abr88,wat01}. The DISKPBB model yields acceptable fits to the spectra of the source in C1, XM3, and XM6 data (see Table 4). The $p$ parameters were found to be consistent with the slim disk model with one exception (C1 data, $p \sim$ 0.75). This may be due to the decrease in the flux of X-6 in C1 data. The inner disk temperatures obtained from DISKPBB fits are consistent with the model \citep{pou07} and similar to some other ULXs (e.g. \citealt{vie06}; \citealt{gla09}; \citealt{sor15}). 

It is possible to constrain the mass of the compact object using the parameters of the disk models (DISKBB and DISKPBB). For this calculation, we used the technique described by \citet{sor15}. Since the spectrum of X-6 is relatively better modeled with DISKBB model in XM3 data, the DISKBB normalization parameter obtained using that observation was adopted for this calculation. We found an apparent inner disk radius of $r_{\mathrm{in}}\sqrt{\cos i} \approx 66$ km. The apparent radius was corrected to the true value using the equation $R_{\mathrm{in}}=\xi \cdot \kappa^{2} \cdot r_{\mathrm{in}}$ equation, where the correction factor $\xi = 0.412$ and $\kappa$ is a spectral hardening factor (see \citealt{kub98}). Assuming $\kappa = 1.7$ \citep{shi95} and disk inclination $i = 60$, the true inner disk radius was calculated as $R_{\mathrm{in}} \approx 100$ km. Using the relation between inner disk radius and mass \citep{mak00}, we found a BH mass of $M \sim 10M_{\sun}$ for a non-spinning BH.

Also, if we consider the normalization parameter of the DISKPBB model in XM3 data, we may calculate another mass value for the compact object. We derived a true inner disk radius of $R_{\mathrm{in}}\sqrt{\cos i} \approx 40$ km using the correction factor $\xi=0.353$ and a spectral hardening factor $\kappa =$ 3 \citep{vie08}. Assuming a moderate disk inclination $i=60$ and taking the mass correction factor as minimum ($\sim$ 1.2) the mass of the compact object in X-6 can be calculated as $M \sim 10M_{\sun}$ for a non-spining BH. This value is consistent with the estimation in the paragraph above.

Three optical counterpart candidates were identified after the astrometric correction. We calculated the X-ray to optical flux ratios for the three sources. This ratio is given as log($f_{\mathrm{X}}$/$f_{V}$)$=$log$f_{\mathrm{X}}$ $+$ $m_{V}$/2.5 $+$ 5.37 where $m_{V}$ is the extinction-corrected visual magnitude and $f_{\mathrm{X}}$ is the unabsorbed X-ray flux in the energy band 0.3$-$3.5 keV \citep{mac82}. Simultaneous X-ray and optical observations are not available for X-6. Therefore we calculated log($f_{\mathrm{X}}$/$f_{V}$) using the minimum (XM5) and maximum (XM6) $f_{\mathrm{X}}$ values adopted from PL model parameters. For source 2, log($f_{\mathrm{X}}$/$f_{V}$) was found to be 1.5$-$1.8. Although these values are within the given ratios for active galactic nuclei (AGNs) ($-$1 to 1.7, \citealt{mac88}), they are acceptable values also for a high-mass X-ray binary. For source 1 and 3, the ratios were found to be 2.2$-$2.5 and 2.0$-$2.3, respectively. These ratios are greater than the values for AGNs, normal stars, normal galaxies, and BL Lac objects \citep{mac88,sto91} and are similar to those for the optical counterparts of other ULXs \citep{fen08,yan11,tao11,avd16}. 

Colors and absolute magnitudes of these sources were obtained (see Table 5). These sources have $M_{V}$ values that are consistent with the optical counterparts of ULXs in Table 4 of \cite{tao11}, which lie in the range $-$7 $<$ $M_{V}$ $<$ $-$3. If we assume that the optical emission of X-6 is dominated by the companion star and we use the Schmidt--Kaler table (\citealt{all82}) of intrinsic colors, then the probable spectral types of sources 1, 2, and 3 can be estimated to be B2$-$A3, B6$-$A5, and B0$-$A7 supergiants, respectively. We also found, by using the {\it HST}/ACS/WFC images, that the counterpart candidates of X-6 possibly belong to a star cluster. After obtaining CMDs for the stars in the cluster and using the Padova isochrones, the ages of sources 1, 2, and 3 have been estimated to be in the ranges 15$-$30, 12$-$14, and 6$-$25 MYr, respectively. The masses of the counterpart candidates are estimated from the PARSEC isochrones by taking into account their ages and absolute magnitudes as 9$-$13 M$_{\sun}$ for source 1, 14$-$16 M$_{\sun}$ for source 2, and 10$-$25 M$_{\sun}$ for source 3. The mass ranges of the candidates are compatible with the donor stars of other ULXs \citep{pat08}.

On the other hand, the optical emission could also arise from the accretion disk. When the disk is dominant, this emission is expected to vary significantly, as is found for most ULXs, e.g., M101 ULX-1 (\citealt{tao11}). If the optical variability of the counterpart candidate is taken into account, source 1 is a promising candidate in the present case and the contribution from the disk cannot be ignored. Therefore, the donor star and the accretion disk may give comparable contributions to the optical emission of X-6.

In summary, even though the source does not exhibit variations in its X-ray spectrum similar to the Galactic BHBs, there are some spectral changes in both HR and spectral model parameters. Nonetheless, it is hard to say anything conclusively about the accretion regime. Additional X-ray data with better statistical quality can constrain the physical parameters better, shedding more light on the origin of the X-ray emission from X-6. Both broadband photometric and high-resolution spectroscopic observations will help to distinguish the optical counterpart and find the origin of the optical emission.

\acknowledgments

The authors thank the anonymous referee for the helpful suggestions that improved the manuscript. We also would like to thank S. Fabrika for his useful comments. This research was supported by the Scientific and Technological Research Council of Turkey (TUBITAK) through project number 113F039. This research is also supported by Cukurova University Research Fund through project number FEF2013D38 and FDK-2014-1998. We thank to TUBITAK for a partial support in using RTT-150 (Russian--Turkish 1.5-m telescope in Antalya) with project number 14ARTT150-571.

\end{document}